\documentclass[journal]{IEEEtran}
\IEEEoverridecommandlockouts

\usepackage{siunitx}
\usepackage{cite}
\usepackage{todonotes}
\usepackage{amsmath,amssymb,amsfonts}
\usepackage{cleveref}
\usepackage{mathtools} 
\usepackage{algorithm,algcompatible}
\usepackage{graphicx}
\graphicspath{{./figures/}}
\usepackage[caption=false,font=normalsize,labelfont=rm,textfont=rm]{subfig}
\usepackage{tikz}
\usepackage{array}
\usepackage{fixltx2e}
\usepackage{url}
\usepackage{booktabs}
\usepackage{float}

\begin{document}

\title{Semi-Supervised First-Person Activity Recognition in Body-Worn Video}
\author{Honglin~Chen,~Hao~Li,~Alexander~Song,~Matt~Haberland,~Osman~Akar,~Adam~Dhillon,~Tiankuang~Zhou, ~Andrea~L.~Bertozzi,~and~P.~Jeffrey~Brantingham%
\thanks{Manuscript created \today. This work was supported by NIJ grant 2014-R2-CX-0101, NSF grant DMS-1737770, and NSF grant DMS-1417674. Tiankuang Zhou was supported by the UCLA-CSST program. The first two authors contributed equally to this work.}%
\thanks{H. Chen (chenhonglin@g.ucla.edu), H. Li (lihao0809@math.ucla.edu), O. Akar (osmanakar1123@hotmail.com ), A. Bertozzi (bertozzi@math.ucla.edu), and J. Brantinghan (branting@ucla.edu) are with the University of California, Los Angeles.}%
\thanks{A. Song (alexandersong@umail.ucsb.edu) is with the University of California, Santa Barbara.}
\thanks{M. Haberland (mhaberla@calpoly.edu) is with California Polytechnic State University.}%
\thanks{T. Zhou (tiankuangzhou@gmail.com) is with the University of Science and Technology of China.}}

\maketitle

\begin{abstract} 
Body-worn cameras are now commonly used for logging daily life, sports, and law enforcement activities, creating a large volume of archived footage. 
This paper studies the problem of classifying frames of footage according to the activity of the camera-wearer with an emphasis on application to real-world police body-worn video. Real-world datasets pose a different set of challenges from existing egocentric vision datasets: the amount of footage of different activities is unbalanced, the data contains personally identifiable information, and in practice it is difficult to provide substantial training footage for a supervised approach. We address these challenges by extracting features based exclusively on motion information then segmenting the video footage using a semi-supervised classification algorithm. On publicly available datasets, our method achieves results comparable to, if not better than, supervised and/or deep learning methods using a fraction of the training data. It also shows promising results on real-world police body-worn video.
\end{abstract}

\section{Introduction}
\IEEEPARstart{W}{ith} the development of body-worn camera technology, it is now possible and convenient to record continuously for a long period of time, enabling video capture of entire days. This technology has been been adopted by law enforcement agencies to log all police activity. The footage serves as a source of evidence and a way to promote transparency and accountability in law enforcement~\cite{bwv-official}. However, this generates a massive amount of first-person footage on a daily basis that is impossible for humans for review. It is hence necessary to develop algorithms to aid in summarizing and indexing such video footage. In this paper, we are concerned with recognizing what the camera-wearer is doing and indexing the footage according to this ``ego-activity''. 

Classifying ego-activities in body-worn video footage has often been studied in the context of sports videos \cite{kitani2011fast} and life-log videos \cite{singh2017trajectory, pirsiavash2012detecting, poleg2016compact, fathi2012learning}. Other than \cite{egomotionclassification}, no experiments have been conducted on real-world police body-worn video. In the present paper, we fill this gap by studying a police body-worn video dataset provided by the Los Angeles Police Department (LAPD) and designing an ego-activity recognition algorithm to handle the unique challenges present in the field dataset. As pointed out in \cite{egomotionclassification}, the real-world human behaviors captured in the footage are diverse; we also observe that the amount of footage of each activity is distributed unevenly. In the provided dataset, the majority of the footage is recorded when the police officers are standing still, walking, or inside a vehicle, but we have also noticed a variety of relatively rare activities such as riding a bike or motorcycle. One reason that footage of an activity is scarce is that the activity is not recorded very often. For instance, we have only found police officers riding a motorcycle in two of the hundred videos we have studied. On the other hand, some activities occur frequently, but each occurrence lasts a very short period of time, e.g. a burst of running in chase of a suspect. It is important for us to correctly recognize these rare activities because often they are the more significant ones to identify. 

Most ego-activity recognition methods require a substantial amount of training footage; supervision is either used to determine the importance of extracted low-level features in a bottom-up system (e.g. \cite{poleg2014temporal}) or used to learn features in a deep-learning approach (e.g. \cite{poleg2016compact}). In the present paper, we consider a semi-supervised approach, in which we utilize a much smaller amount of labeled training data than a typical supervised method. Although body-worn video footage is abundant, the majority of the footage is recording routine activities. This makes it tedious and difficult to locate and annotate a large amount footage of rare but significant activities. In the application to police body-worn video, which is highly sensitive because it contains personally identifiable information, the annotation process is even more time-consuming due to strict data security protocols. 

The proposed semi-supervised approach is based on similarity graphs. It first quantifies the similarities between pairs of data points, i.e. short pieces of video, according to handcrafted, motion-based features adapted from \cite{ryoo2013first}. Then, it spreads the label information from a small set of manually labeled fidelity points to unlabeled data. We propose the use of handcrafted features instead of deep-learning features so that we can ensure that the features do not compromise personally identifiable information in police body-worn videos. For the same data security reason, we employ features based exclusively on motion cues without object detection and tracking. With the aid of the Nystr\"{o}m extension, the graph-based semi-supervised classification method is scalable to handle the enormous size of body-worn video datasets. 

Besides the police body-worn video dataset, we also benchmark the proposed method on publicly available datasets and demonstrate its comparable performance to supervised methods although it only uses a fraction of training data. 

The paper is organized as follows. In \cref{sec:relatedworks} we survey related work on analyzing egocentric vision and activity recognition. In \cref{sec:feature}, we introduce our feature extraction and semi-supervised learning method. We report our experimental setup and results in \cref{sec:exp}. Finally, the conclusions and future work follow in \cref{sec:conclusion}. 

\section{Related Works}\label{sec:relatedworks}

Research in summarizing and segmenting egocentric videos recorded by body-worn cameras dates back to the early 2000s \cite{aizawa2001summarizing}. Since then, this has been an active research area due to the advancement in computer vision, machine learning, and deep learning~\cite{del2017summarization}; here we review work most relevant to our own. 

The task of activity recognition in body-worn video can be categorized into three lines of research: (1) one relies on object-hand interactions and video content (i.e. what objects and people are in the video), (2) one uses the motion of the camera, and (3) ones uses a combination of the previous two. Typically, neither object-hand interactions nor the motion of the camera is directly available as metadata in egocentric vision datasets, so all three lines of research start with inferring respective pieces of information from raw video footage. 

Works following the first approach rely on object detection and tracking to classify the camera-wearer's activities, for instance, \cite{fathi2011understanding,fathi2012learning,pirsiavash2012detecting,li2015delving,spriggs2009temporal}. Popular benchmark datasets used to validate methods focusing on hand-object interactions are the GTEA and GTEA Gaze+ datasets, provided by \cite{fathi2012learning}, and ADL-short in \cite{singh2017trajectory} and ADL-long in \cite{pirsiavash2012detecting}. The GTEA and GTEA Gaze+ datasets are recorded by Tobii eye-tracking glasses when wearers are cooking in a natural setting, so these two datasets contain eye-gaze direction information not typically available in other body-worn video datasets. Both ADL dataset are recorded with a chest-mounted camera when the wearers are performing various daily tasks indoors. The aforementioned datasets are different from our police body-worn video dataset, which was recorded outdoor and usually did not capture police officers' hand movement, so we do not pursue this thread of research. 

The second line of research is to recognize activities based on motion analysis. A wide variety of motion features have been proposed in the literature. \cite{kitani2011fast} uses a histogram-based motion feature to classify sports activities in videos recorded by head-mounted GoPro cameras. \cite{ryoo2013first} proposes a motion descriptor that inspired our feature selection method.  \cite{egomotionclassification} uses inferred camera movement signals and their dominant frequencies. Many ways of incorporating temporal information in motion analysis are proposed; for instance,  \cite{ryoo2015pooled} proposes to apply multiple temporal pooling operators to any per-frame motion descriptor. Deep convolutional neural networks are also used to extract motion features; for instance, \cite{abebe2017long} learns a motion representation by using 2D convolution neural network on stacked spectrograms and a Long Short-Term Memory (LSTM) network. With multiple available features extracted, \cite{ozkan2017boosted} proposes a multiple kernel learning method to combine local and global motion features.  A benchmark dataset for this line of research is the HUJI EgoSeg dataset provided by \cite{poleg2014temporal}, which is recorded when the wearer is performing a variety of activities in both indoor and outdoor settings. As in the HUJI EgoSeg dataset, we observe that many activities of interest in our police body-worn video dataset induce distinctive camera movement patterns, and so we focus on a motion-based approach. The proposed approach differs from the aforementioned methods in that it is semi-supervised; we demonstrate in section \ref{sec:huji} that it achieves comparable performance to the supervised methods on the HUJI EgoSeg dataset using a fraction of training data.  

For the third line of research, methods that utilize both appearance (i.e. object recognition and tracking) and motion cues are often combined with  deep learning. 
Both \cite{poleg2016compact} and \cite{ma2016going} use a two-stream deep convolution neural network, one stream for images and another stream for optical flow fields, to discover long-term activities in body-worn video. Both \cite{bhatnagarunsupervised} and \cite{wang2018deep} use an auto-encoder network to extract motion and appearance features in an unsupervised fashion. We note that features extracted from appearance cues using a convolutional neural network may be used to reconstruct the original frame, which can potentially be used to recover personally identifiable information, so we do not pursue this line of approach. 

In \cite{egomotionclassification}, the authors also study ego-activity recognition in police body-worn video. We improve upon their work by choosing a more sophisticated feature than theirs to handle the increased diversity of activities in our much larger dataset. We demonstrate the improved performance of the proposed method in section \ref{sec:exp}. To the best of our knowledge, no other experiment results on real-world police body-worn video have been reported in the literature.

\begin{figure*}
\includegraphics[width=1\textwidth]{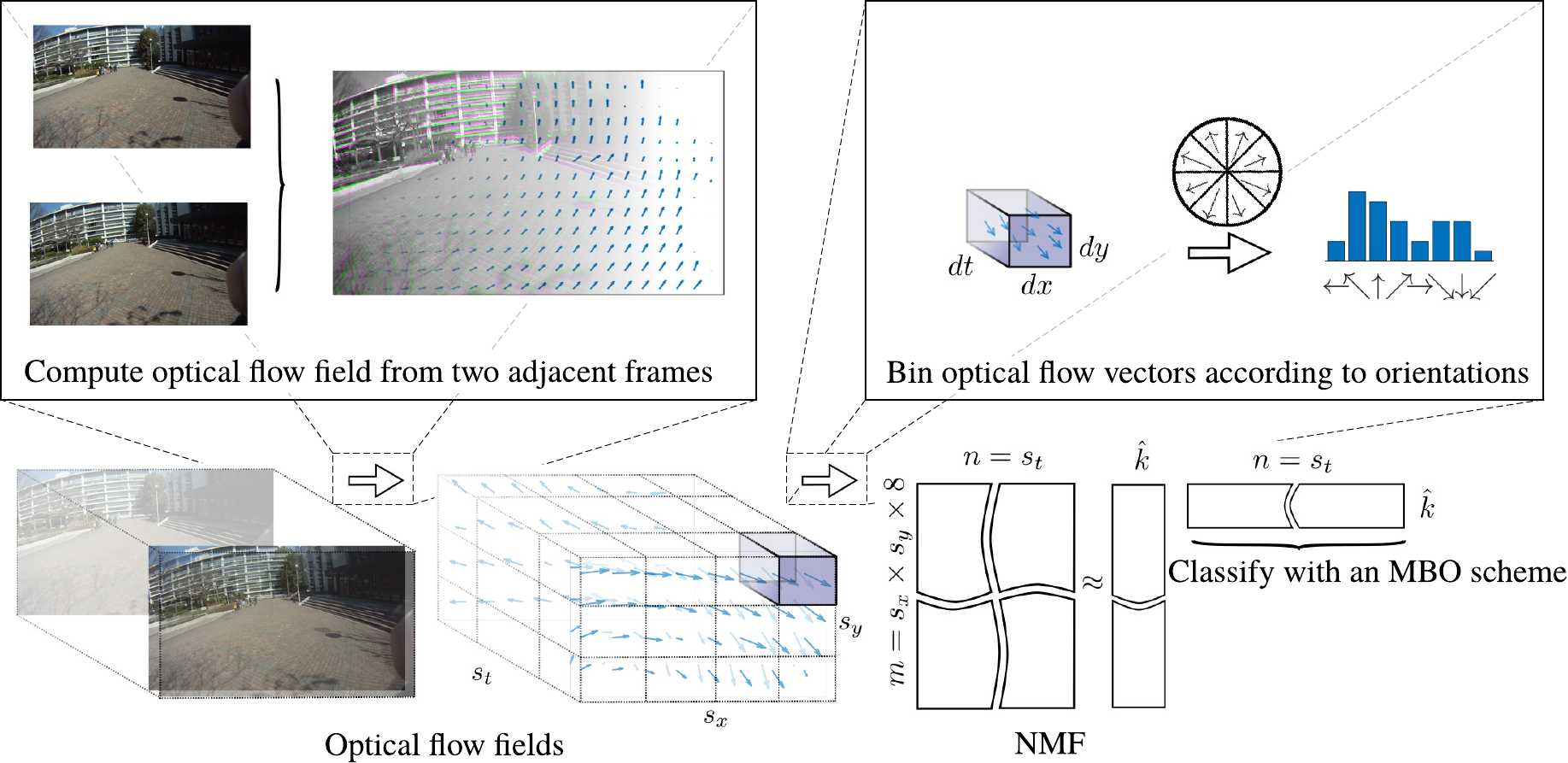}
\caption{A summary of the proposed method. First, we compute a dense optical flow field for each pair of consecutive frames. We then divide each optical flow field into $s_x\times s_y$ spatial regions, where each region consists of $dx\times dy$ pixels, and divide the video into $s_t$ temporal segments, where each segment consists of $dt$ frames. For each $dx \times dy \times dt$ cuboid, we count the number of flow vectors with direction lying within in each octant, yielding a $s_x\times s_y$ histogram for each segment of video. We reshape and concatenate each histogram into a single feature vector of dimension $s_x \times s_y \times 8$ describing the motion that occurs within the video segment. The dimension of the feature vectors is reduced with NMF and we smooth them with a moving-window average operator. Finally, we classify the smoothed features with a semi-supervised MBO scheme.    \label{fig:flowchart}}\end{figure*}

\section{Method}\label{sec:feature}
We start with extracting features based on motion cues from the video. The extracted motion features are potentially high-dimensional, so they are compressed to a lower-dimensional representation to alleviate computational burden. Finally, we classify the video footage with the low-dimensional representation using a graph-based semi-supervised learning method that only requires 10\% training data from each class of activity. The flowchart \cref{fig:flowchart} summarizes the proposed system, which we detail below.

\subsection{Motion Descriptor}
Our motion descriptor is similar to the one presented in \cite{ryoo2013first} except for the final dimension reduction step: \cite{ryoo2013first} uses the principle component analysis (PCA) whereas we choose the non-negative matrix factorization (NMF) because the features are inherently non-negative. Before we compute any feature, we resize all video frames to have a resolution of $576 \times 1024$ and hence an aspect ratio of $16:9$, allowing us to choose a uniform set of video parameters across all datasets.  

\subsubsection{Dense Optical Flow Fields}
Dense optical flow fields \cite{horn1981determining,lucas1981iterative,barron1994performance,farneback2003two}, which describe relative motion between objects in the scene and the camera, form the basis of our motion analysis.  Optical flow fields are fields of two-dimensional vectors defined on the two-dimensional domain of images. In the discrete setting, an optical flow field associates each pixel in an image with an optical flow vector which consists of a horizontal and vertical component. An optical flow field is calculated from a pair of consecutive frames under the assumption that pixels displaced according to the optical flow field should preserve their intensities after the displacement. Assuming that the objects recorded in a pair of frames are static, the optical flow field encodes the movement of the camera and hence the movement of the camera-wearer. Although this assumption does not necessarily hold perfectly for real-world body-worn video footage, static background objects often cover the majority of frames, and thus we can use optical flow fields to estimate the movement of the camera-wearer. Even when this assumption is not true, we have found that optical flow fields induced by the movement of objects instead of the camera-wearer are still helpful in certain situations. For instance, they characterize driving a car by the static interior of the vehicle and the movement in the windshield region. This is also observed in the experiments conducted by \cite{poleg2016compact}; the authors find distinctive patterns of optical flow fields in the windshield region that correspond well to the camera-wearer driving a car.   

\subsubsection{Histograms on Dense Optical Flow Fields}
\begin{algorithm}
\caption{ Global Motion Descriptor}
\begin{algorithmic}[1]
\STATE \textbf{Input:} Optical flow fields matrix $O\in\mathbb{R}^{n_f\times n_x\times n_y \times 2}$
\STATE \textbf{Output:} Matrix $X\in\mathbb{R}^{s_t\times (s_x \cdot s_y \cdot 8)}$
\STATE \textbf{Initialize} $dt=60, dx=dy=64, s_x=\frac{n_x}{dx}, s_y=\frac{n_y}{dy}, \newline s_t=\lfloor \frac{n_f}{dt}\rfloor$, histogram count matrix $C\in\mathbb{R}^{s_t\times s_x \times s_y \times 8}$

\FOR{$i = 0:s_t$}
\FOR{$j = 0:s_x$}
\FOR{$k = 0:s_y$}

\STATE \% Step 1. Partition: 
\STATE cuboid = $O[idt:(i\text{+}1)dt, jdx:(j\text{+}1)dx,\newline {\qquad \qquad \qquad \qquad}kdy:(k\text{+}1)dy,:]$
\STATE \% reshape: $\mathbb{R}^{dt\times dx\times dy \times 2} \mapsto \mathbb{R}^{(dt\cdot dx\cdot dy) \times 2}$
\STATE cuboid = reshape(cuboid)
\STATE \% Step 2. Histogram count: 
\FOR{$l = 0,1,\cdots,(dt\cdot dx\cdot dy)$}
\STATE $v$ = cuboid$[l,:]$
\STATE $\theta$ = phase($v$)
\STATE $bin$ = $\lfloor \theta/\frac{\pi}{4} \rfloor$
\STATE $C[i,j,k,bin]$++
\ENDFOR
\ENDFOR
\ENDFOR	
\ENDFOR
\STATE \% reshape: $\mathbb{R}^{s_t\times s_x \times s_y \times 8} \mapsto\mathbb{R}^{(s_x \cdot s_y \cdot 8)\times s_t }$
\STATE $X$ = reshape($C$) 
\end{algorithmic}
\end{algorithm}
Using optical flow fields is common in classifying ego-activities. Different motion features are effectively different ways of aggregating them. For instance, authors of \cite{kitani2011fast,ryoo2013first,ryoo2015pooled} bin optical flow vectors to construct features in the form of concatenated histograms, \cite{poleg2016compact,wang2018deep,bhatnagarunsupervised} aggregate them via convolution kernels, and \cite{egomotionclassification,poleg2014temporal} infer camera movement using unaggregated optical flow fields as input. In our case,
we compute the motion descriptors, proposed in \cite{ryoo2013first}, as histograms of extracted dense optical flow vectors. We bin the vectors according to their locations in the frames and orientations, and then count the number of vectors in each bin. Note that we lose magnitude information in this process because the bins only correspond with locations and orientations. 
The features proposed in \cite{kitani2011fast} retain magnitude information by further grouping optical flow vectors 
according to their magnitudes, but in our experiments we observe comparable performance using the simplified features. 

To compute the motion descriptors from the optical flow fields, we consider a video as a 3D volume with frames (optical flow fields) stacked along the time axis. We spatially divide each frame into $s_x$ by $s_y$ rectangular regions of fixed width $dx$ and height $dy$ pixels; the choice of $dx$ and $dy$ determines the spatial resolution of the final features. We have found that choosing $dx$ and $dy$ that are divisible by the total number of pixels in length and height, respectively --- yielding $s_x = 16$ and $s_y = 9$ --- gives good performance on all datasets tested. We also divide the video into $s_t$ video segments, each with a fixed time duration $\Delta T$, that is, $dt$ frames. We choose $\Delta T$ depending on the time scale of the ego-activities that we wish to classify. For instance, we choose $\Delta T = 0.2$ second for videos containing a mix of long term and short term ego-activities, whereas we choose $\Delta T = 4$ seconds if we wish to classify relatively long-term activities. The choice of $\Delta T$ also determines the computation cost of the subsequent analysis. A finer time resolution, i.e. a smaller $\Delta T$, yields more video segments for a given video and hence results in more computations. 

Consider the optical flow vectors in each $dx \times dy \times dt$ volume. We place each of them into one of the pre-defined eight histogram bins based on its orientation. Formally, a vector with a directional angle $\theta$ is placed in bin $\left \lfloor{\theta/\frac{\pi}{4}}\right \rfloor$. Repeating the above steps for every $dx \times dy \times dt$ volumes in each video segment of duration $\Delta T$, we obtain a feature vector with a dimension of $s_x\times s_y\times 8$ for each segment, which we reshape into a single column vector. By repeating the above procedures for every video segments of length $\Delta T$ and stacking obtained feature vectors, we obtain a data matrix $X$ with the number of columns equal to the number of segments in the video.

\subsubsection{Non-negative Matrix Factorization}
The concatenated histograms for each video segment can have $9\times 16\times 8= 1152$ entries, which can potentially be expensive to compute with. To alleviate this problem, we employ dimension reduction techniques. In \cite{ryoo2013first}, the authors use the principal component analysis (PCA) to perform dimension reduction. However, we use non-negative matrix factorization (NMF) \cite{lee2001algorithms} because the concatenated histograms are inherently non-negative. NMF is widely used in the context of topic modeling, where users want to learn topics, a collection of words that often co-occur in textual documents, each of which is represented by a histogram of words. In our case, each video segment is represented by a histogram of ``motion words''; each motion word is the movement of a specific orientation in a specific region of the frame. Analogously, a topic --- a collection of motion words --- describes a global movement pattern. We then model the concatenated histogram of motion words of each video segment as a non-negative linear combination of the topics. 

NMF factorizes a non-negative $m\times n$ matrix $X$ (in our case, $m = s_x\times s_y\times 8$ and $n = s_t$) into the product of two low rank non-negative  $m\times \hat{k}$ and $\hat{k} \times n$ matrices $V$ and $H$. The number $\hat{k}$ is chosen by the users according to their computation resources and tuned based on the resulting performance. We have found that $\hat{k} = 50$ works well for all considered datasets. Formally, this is achieved by solving the following constrained minimization problem,
\begin{equation}
\min_{V,H} \| X - VH \|_F^2, \text{subject to}\ V \geq 0, H\geq 0,\label{nmf}
\end{equation}
where $\|\cdot\|_F$ denotes the Frobenius norm. Each column in $V$ represents a basis vector (a topic), and each entry in $H$ represents the non-negative linear combination coefficients. Each column in the matrix $H$ is the feature vector for a single video segment, which will be passed into our classification algorithm after a post-processing step (detailed in \cref{sec:postprocessing}).

We also note that we do not necessarily need to perform NMF every time we obtain a new video. We may choose to fix $V$ which we obtain by applying NMF to the initial dataset. Then we only need to compute the combination coefficients $H^\mathrm{new}$ for the new videos $X^\mathrm{new}$ by solving a non-negative least squares problem 
\begin{equation}
H^{\mathrm{new}} = \arg\min_{H} \|X^\mathrm{new} - VH\|_F^2,
\end{equation}
which can be solved very efficiently using methods such as ones proposed in \cite{lawson1995solving}.

 The dimension reduction step also helps to secure personally identifiable information. Since we do not make use of $V$ in the classification algorithm, there is no need to save it. Without the basis, it is impossible to reconstruct the data matrix $X$ and let alone the content of the videos. 

\subsubsection{Post-processing}\label{sec:postprocessing}
We assume a certain degree of temporal regularity of the extracted features: the duration of activities is typically much longer than transitions between them, and so transitions are relatively rare.  We note that none of our previous feature extraction procedure takes advantage of this temporal regularity. Each optical flow field is computed from only two adjacent frames, motion descriptors are aggregated within non-overlapping video segments, and NMF treats columns in the data matrix $X$ (motion descriptors of video segments) independently. Methods exploiting temporal regularity have been proposed before. In \cite{ryoo2015pooled}, for instance, the authors apply multiple temporal pooling operators to the extracted per-frame motion and visual features and use the outputs as additional features. We choose a simpler approach, in which we apply a single moving-window average operator on each row of $H$ and then pass these averaged features to the classification method. We determine the window size of the moving-window average operator experimentally for each dataset. Choosing a large window size may eliminate distinct features of short-term activities, so the choice depends on the types of activities in the dataset as well as the chosen value of $\Delta T$. 

\subsection{Classification Method}
\label{sec:class}
\subsubsection{
Graph-based semi-supervised classification method}
\label{MBO_CLASS}
Recently, graph-based semi-supervised and unsupervised learning methods have been successfully applied to image processing \cite{merkurjev2013mbo} and classification of high-dimensional data such as hyperspectral images \cite{merkurjev2014graph, gIyer,zhu2017unsupervised,meng2017hyperspectral} and body-worn videos \cite{egomotionclassification}. In this section, we outline one of these methods based on minimizing the graph Total Variation, which has been studied in \cite{bertozzi2016diffuse,garcia2014multiclass,merkurjev2014diffuse}.
We consider each data point (i.e. video segment) as a node in a weighted graph. The edge weight between a pair of nodes $i$ and $j$ is given by the similarity
\begin{equation}
w_{ij} = \exp\left(-\frac{\|H_i - H_j\|_2^2}{\tau_{ij}}\right),\label{eq:weight}
\end{equation}
where $\|\cdot\|_2$ denotes the 2-norm of a vector and $\tau_{ij}$'s are scaling constants. Here $H_i$ is the $i$th column of matrix $H$ obtained from NMF. The scaling constants can either be the same chosen $\tau$ for all pairs of $i$ and $j$, or chosen locally for each individual pair \cite{zelnik2005self}. We choose the local scaling constants $\tau_{ij}=\tau_i \tau_j$ where $\tau_i$ is the distance between $i$ and its $K$th nearest neighbor.

We aim to partition $n$ nodes into $c$ classes (i.e. ego-activities) such that 
\begin{enumerate}
\item similar nodes between which edge weights are large (i.e. $w_{ij}$'s are close to 1 should be in the same class, and
\item fidelity nodes (i.e. manually labeled nodes) should be classified according to their labels. 
\end{enumerate}
To achieve 1), we optimize the graph Total Variation (TV) defined as follows. Let $u$ be an $\{0,1\}^{c}$-valued assignment function on the set of nodes, that is $u_\ell(i) = 1$ meaning we assign the $i$th data point to class $\ell$. We can then define the graph Total Variation 
\begin{equation}
|u|_{TV} =\frac{1}{2} \sum_{i,j=1}^n w_{ij} \|u(i) - u(j)\|_1\label{eq:tv}
\end{equation}

We observe that \eqref{eq:tv} admits a trivial minimizer that is constant across all nodes. To avoid this problem and to incorporate the fidelity data, we introduce a quadratic data fidelity term
\begin{equation}
F(u) =  \frac{1}{2} \sum_{i=1}^nM(i)\|u(i) - f(i)\|_2^2\,,
\end{equation}
where $M(i) = 1$ if node $i$ is chosen as fidelity and 0 otherwise, and $f(i)\in \{0,1\}^{c}$ encodes the known label of node $i$. We weight the fidelity term by a positive parameter $\eta$ to balance the graph TV term and the fidelity term in the objective function,
\begin{equation}
\frac{1}{2}|u|_{TV} + \eta F(u). \label{eq:objective}
\end{equation}

Instead of minimizing \eqref{eq:objective} directly, which is discrete and combinatorial, we solve the Ginzburg-Landau relaxation \cite{bertozzi2016diffuse} for $u(i)\in \mathbb{R}^{c}$. Namely, we replace the graph Total Variation $|u|_{TV}$ with 
\begin{equation}
GL_\epsilon(u) = \frac{1}{4}\sum_{i,j=1}^n w_{ij} \|u(i)-u(j)\|_2^2 + \frac{1}{\epsilon}\sum_{i=1}^nP\left(u(i)\right),
\label{eq:gl}
\end{equation}
where $\epsilon$ is a small positive constant, and $P$ is a multi-well potential with minima at the corners of the unit simplex, for instance
\begin{equation}
P\left(u(i)\right) = \prod_{\ell = 1}^{c}\frac{1}{4}\|u(i) - e_\ell\|_2^2,
\end{equation}
where $e_\ell$ is the unit vector in $\mathbb{R}^{c}$ in the $\ell$th direction. The authors of \cite{van2012gamma} prove the following $\Gamma$-convergence
\begin{equation}
GL_\epsilon(u) \xrightarrow{\Gamma} \begin{cases}
|u|_{TV} & \text{if $u$ is binary}\\
+\infty & \text{otherwise}
\end{cases}\label{eq:gammaconv}
\end{equation}
as $\epsilon\rightarrow 0$ in the case of $c = 2$. The $\Gamma$-convergence ensures that the minimizers of $GL_\epsilon(u)$ approach the minimizers of $|u|_{TV}$.

After the Ginzburg-Landau relaxation, we arrive at the objective function
\begin{equation}
GL_\epsilon(u) + \eta F(u),\label{eq:globj}
\end{equation}
which we minimize with respect to $u$.

To formulate \eqref{eq:globj} in terms of matrices, we first identify $u$ and $f$ by a $n\times c$ matrix where $u_{i\ell} = u_\ell(i)$ and $f_{i\ell} = f_\ell(i)$. We let $W$ be the matrix of $w_{ij}$'s, and $D$ be an $n\times n$ diagonal matrix with the $i$th entry $d_i$ being the strength of node $i$, i.e. $d_i = \sum_{j=1}^n w_{ij}$, and then define the graph Laplacian 
\begin{equation}
L = D- W.
\end{equation}
We also let $M$ be an $n\times n$ diagonal matrix of which the $i$th entry is $M(i)$ indicating whether node $i$ is chosen as fidelity. If we define $L$ and $M$ this way, we 
can write \eqref{eq:globj} in the matrix form
\begin{equation}
\frac{1}{2}\mathrm{trace}\left(u^T L u\right) + \frac{1}{\epsilon}\sum_{i=1}^n P\left(u_i\right) + \frac{\eta}{2}\|M(u - f)\|_2^2.\label{eq:matrix}
\end{equation}
In graph clustering, unsupervised learning, and community detection literature, the graph Laplacian is often normalized to guarantee convergence to a continuum differential operator with a large number of data points (see, for instance, \cite{bertozzi2016diffuse}). One popular version of normalized graph Laplacian is the symmetric Laplacian
\[
L_s = I - D^{-\frac{1}{2}}W D^{-\frac{1}{2}}.
\]
If we substitute $L$ for $L_s$, the first quadratic term of \eqref{eq:matrix} becomes
\[
\mathrm{trace}\left(u^TL_su\right) = \frac{1}{2}\sum_{i,j=1}^n w_{ij}\left\Vert\frac{u(i)}{\sqrt{d_i}} - \frac{u(j)}{\sqrt{d_j}}\right\Vert_2^2.
\]

The methods described in the remainder of this paper carry over regardless of which graph Laplacian is used, and the notation $L$ is a placeholder for any choice of graph Laplacian. In our experiments, we choose to use the symmetric graph Laplacian $L_s$ because it permits the use of efficient and simple computation routines to approximate its eigenvalues and eigenvectors.
\subsubsection{Optimization scheme}
Minimizing \eqref{eq:matrix} using the standard gradient descent method yields
\begin{equation}
\frac{\partial u}{\partial t} = - Lu - \frac{1}{\epsilon}\nabla \hat{P}(u) - \eta M(u-f),\label{eq:grad}
\end{equation}
where $\hat{P}(u) = \sum_{i=1}^n P(u_i)$. The steady-state solution of \eqref{eq:grad} is a stationary point of \eqref{eq:matrix}. 
This is known as the graph Allen-Cahn equation, and 
we refer readers to \cite{luo2017convergence}
for a convergence analysis of the graph Allen-Cahn scheme. We follow \cite{meng2016openmp} to use a variant of the MBO (Merriman-Bence-Osher) scheme to approximate and solve \eqref{eq:grad}. In short, we first randomly initialize $u^0$, which we use as the initial condition for \eqref{eq:grad}. We then alternate between the following two steps:
\begin{enumerate}
\item [1.] \emph{Diffusion}: for given $u^k$, we obtain $u^{k+\frac{1}{2}}$ by solving a force-driven heat equation
\begin{equation}
\frac{\partial u}{\partial t} = -L u - \eta M(u-f),\label{eq:heat}
\end{equation}
for $t_k \le t \le t_k + \frac{1}{2}\Delta t$, where $\Delta t$ is a parameter. 
\item[2.] \emph{Threshold}: we threshold $u^{k+{\frac{1}{2}}}$ to obtain $u^{k+1}$, i.e.
\begin{equation}
u^{k+1}(i) = e_\ell \;\text{, where } \ell = \arg\max_{\hat{\ell}} u^{k+\frac{1}{2}}_{i\hat{\ell}}.\label{eq:threshold2}
\end{equation}
For a small $\epsilon$, this approximates solving
\begin{equation}
\frac{\partial u}{\partial t} = - \frac{1}{\epsilon}\nabla \hat{P}(u)\label{eq:threshold}
\end{equation}
for $t_k + \frac{1}{2}\Delta t \le t \le t_{k+1} = t_k + \Delta t$.
\end{enumerate}

Choosing $\Delta t$ is delicate. If it is too small, $u^{k+1}  = u^{k}$ after thresholding, whereas if it is too large, $u$ converges to the steady-state solution of \eqref{eq:heat}, 
\[
(L+\eta M)^{-1}Mf,
\]
in one diffusion step, independent of the initial condition $u^k$. Either way, extreme $\Delta t$ results in a ``freezing'' scheme. 
In \cite{van2014mean}, the authors give guidance on how to choose $\Delta t$ in the case of unnormalized graph Laplacian, $c = 2$ (i.e. binary classification), and $\eta = 0$. Currently, there is no analogous result for a symmetric graph Laplacian,  multi-class classification, and nonzero $\eta$. We have found, however, that $\Delta t = 0.1$ gives nontrivial dynamics (i.e. convergent and not ``freezing'') on all datasets used in testing. 
\subsubsection{Numerical methods}
We follow \cite{bertozzi2016diffuse, garcia2014multiclass} to employ a semi-implicit ordinary differential equation solver to solve \eqref{eq:heat}, and use a pseudo-spectral method coupled with the Nystr\"{o}m extension to make the ordinary differential equation solver efficient. We note that the graph Laplacian matrix $L$ is large, with $n^2$ entries where $n$ is the number of data points; it is also not inherently sparse, which makes approximation techniques such as the Nystr\"{o}m extension necessary.   

For the ordinary differential equation solver, we take $N_{step}$ time steps to reach $u^{k+\frac{1}{2}}$ from $u^k$, where $N_{step}$ is a parameter to choose. Formally, we let $u^{k,s}, s = 0,1,\cdots, N_{steps}$ denote the numerical solutions of \eqref{eq:heat} at intermediate time $t_k + s\delta t$, where $\delta t = \Delta t/{2 N_{step}}$. We solve
\begin{equation}
\frac{u^{k,s+1} -u^{k,s}}{\delta t} = - Lu^{k,s+1} - \eta M (u^{k,s} - f)
\label{eq:update}
\end{equation}
for $u^{k,s+1}$. We use $N_{step} = 10$ to ensure convergence of the ordinary differential equation solver when $\eta < 500$ and $\Delta t = 0.1$. 

We use a pseudo-spectral method to solve Equation (\ref{eq:update}). We project the solution $u$ onto an orthonormal eigenbasis of the graph Laplacian $L$, or an eigen-subbasis that consists of $N_{eig}$ eigenvectors corresponding to the smallest $N_{eig}$ eigenvalues. We detail how we compute the spectrum of $L$ with the Nystr\"{o}m extension in \cref{sec:eig}. Choosing a modest $N_{eig}\ll n$ will greatly improve the efficiency of the algorithm because solving \eqref{eq:update} only requires $O(nN_{eig})$ operations if the eigenvectors and eigenvalues of $L$ are provided. Suppose $\Phi$ is an $n\times N_{eig}$ eigenvector matrix, of which the $j$th column $\phi_j$ is the eigenvector of $L$ corresponding to the $j$th smallest eigenvalue $\lambda_j$, and $\Lambda$ is the diagonal matrix containing all $N_{eig}$ smallest eigenvalues $\lambda_j$'s. We let $a$ denote the coordinates we obtain by projecting columns of $u$ onto the eigen-subspace spanned by columns of $\Phi$, i.e. $a = \Phi^T u$. Solving \eqref{eq:update} in the eigen-subspace is simply
\begin{eqnarray*}
a^{k,s+1} &=& (I +\delta t \Lambda)^{-1}a^{k,s} - \delta t\cdot \eta\Phi^T M (u^{k,s}-f)\,,\\
u^{k,s+1} &=& \Phi a^{k,s+1}\,.
\end{eqnarray*}

\begin{algorithm}
\caption{ Graph MBO scheme \cite{bertozzi2016diffuse}}
\begin{algorithmic}[1]
\STATE \textbf{Input:} $\Phi, \Lambda, M, f, \eta$, and initial guess $u^0$.
\STATE \textbf{Output:} $u$.
\STATE \textbf{Initialize} $u^{0,0} = u^0, a^{0,0} = \Phi^T u^0$.
\FOR{$k = 1,2,\cdots,$ MaxIter or $u^k$ has converged}
\STATE a. Diffusion:
\FOR{$s = 0,1,\cdots, N_{step}-1$}
\STATE $a^{k,s+1} = (I +\delta t \Lambda)^{-1}a^{k,s} - \delta t\cdot \eta\Phi^T M (u^{k,s}-f)$.
\STATE $u^{k,s+1} = \Phi a^{k,s+1}$. 
\ENDFOR
\STATE b. Threshold $u^{k+1/2}:=u^{k,N_{step}}$ :
\FOR{$i = 1,2,\cdots, n$}
\STATE $u^{k+1,0}(i) = e_{\ell}$, where $\ell = \arg\max_{\hat{\ell}} u_{i\hat{\ell}}^{k, N_{step}}$
\ENDFOR
\ENDFOR
\end{algorithmic}
\end{algorithm}
 
\subsection{Nystr\"{o}m extension}
\label{sec:eig}
We employ the Nystr\"{o}m extension~\cite{fowlkes2004spectral}, which approximates the eigenvectors and eigenvalues of $L$ with $O\left(nN_{eig}^3\right)$ computation complexity and $O(nN_{eig})$ memory requirement.  With $N_{eig}\ll n$, the computation complexity and memory scales linearly with respect to the number of data points. The idea of the Nystr\"{o}m extension is to uniformly randomly sample a smaller set of data points $A \subset \{1,2,\cdots,n\}$ with $|A| = N_{sample} \ll n$, perform spectral decomposition on an $N_{sample}\times N_{sample}$ system calculated from the set of data points $A$, and then interpolate the result to obtain an approximation to the spectral decomposition of the entirety of $L$. Let $B$ be the complement of $A$, i.e. $A\cup B = \{1,2,\cdots,n\}$ and $A\cap B = \emptyset$. Let $W_{AA}$ denote the weights associated with nodes in set $A$, and similarly, let $W_{AB} = W_{BA}^T$ denote weights between nodes in set $A$ and $B$. If we reorder the nodes so that $A = \{1,2,\cdots,N_{sample}\}$ and $B = \{N_{sample}+1,N_{sample}+2,\cdots,n\}$, we can rewrite 
\begin{equation}
W = \begin{bmatrix}
W_{AA} & W_{AB} \\
W_{BA} & W_{BB}
\end{bmatrix}.
\end{equation}
It can be shown \cite{fowlkes2004spectral} that the matrix $W_{BB}$ can be approximated by $W_{BB}\approx W_{BA}W_{AA}^{-1}W_{AB}$ in the context of approximating the spectral decomposition. The Nystr\"{o}m extension uses this property to approximate the spectrum of $W$, and henceforth $L$. We summarize the Nystr\"{o}m extension algorithm to approximate the spectrum of symmetric graph Laplacian in \Cref{alg:nystrom}. An analogous algorithm for unnormalized graph Laplacian can be found in \cite{bertozzi2016diffuse}. In \Cref{alg:nystrom}, $\mathbf{1}$ denotes a vector of one's that is used to compute the strength of each nodes, i.e. the sum of weights, and let $X./Y$ denote component-wise division between two matrices $X$ and $Y$ of the same size. We let $\sqrt{X}$ denote the non-negative square root of each component of any non-negative matrix $X$, and if $X$ is positive definite with the spectral decomposition $X = Q\Gamma Q^T$, we let $X^{1/2} = Q\Gamma^{1/2}Q^T$ and similarly $X^{-1/2} = Q\Gamma^{-1/2}Q^T$.   

\begin{algorithm}
\caption{ Nystr\"{o}m Extension for symmetric graph Laplacian\cite{bertozzi2016diffuse}\cite{fowlkes2004spectral}\label{alg:nystrom}}
\begin{algorithmic}[1]
\STATE \textbf{Input:} $\{H_i\}_{i=1}^n$ and $\{\tau_{ij}\}_{ij=1}^n$.
\STATE \textbf{Output:} $\Phi, \{\lambda_j\}_{j = 1}^{N_{eig}}$.
\STATE Randomly sample $A \subset\{1,2,\cdots,n\}$ with $|A| = N_{sample} \ge N_{eig}$ and $B$ such that $A\cup B = \{1,2,\cdots,n\}$.
\STATE Compute $W_{AA}$ and $W_{AB}$ using \eqref{eq:weight}.
\STATE Compute the strength of nodes in $A$, $d_A = W_{AA}\mathbf{1}$.
\STATE Approximate the strength of nodes in
$B$, $d_B = W_{BA}\mathbf{1} + W_{BA}W_{AA}^{-1}W_{AB}\mathbf{1}$.
\STATE Normalize $W_{AA} = W_{AA}./\sqrt{d_A d_A^T}$.
\STATE Normalize $W_{AB} = W_{AB}./\sqrt{d_A d_B^T}$.
\STATE Perform spectral  decomposition on $W_{AA} + W_{AA}^{-1/2}W_{AB}W_{AB}^TW_{AA}^{-1/2}$ to obtain the $N_{eig}$ largest eigenvalues $\{\xi_i\}_{i=1}^{N_{eig}}$ and the corresponding eigenvectors $\{\psi_{i}\}_{i=1}^{N_{eig}}$. We let $\Psi$ denote the matrix of the eigenvectors and $\Xi$ be a diagonal matrix with $\xi_i$'s on the diagonal.  
\STATE Output $\lambda_i = 1-\xi_i$, and $\Phi = \begin{bmatrix}
W_{AA}^{1/2} \\
W_{BA} W_{AA}^{-1/2}
\end{bmatrix}\Psi \Xi^{-1/2}$.
\end{algorithmic}
\end{algorithm}

\section{Experiments}
\label{sec:exp}
We apply the proposed method on two publicly available datasets, the QUAD dataset \cite{kitani2011fast} and the HUJI EgoSeg dataset \cite{poleg2014temporal}, and compare our results to those reported in \cite{kitani2011fast, ryoo2015pooled,tran2015learning,poleg2016compact,poleg2014temporal}. We also apply both our method and the one proposed in \cite{egomotionclassification}\footnote{with the implementation kindly provided by the authors of \cite{egomotionclassification}} on a police body-worn video dataset provided by the LAPD. Our experimental procedures and parameters are summarized in TABLE \ref{table:para}. The measures of success we use are precision 
\[
\frac{\text{True Positive}}{\text{True Positive + False Positive}}\,
\]
and recall
\[
\frac{\text{True Positive}}{\text{True Positive + False Negative}}\,,
\]
within each class, mean precision and recall directly averaged over all classes, and the overall accuracy, i.e. the percentage of correctly classified data points. 

The feature extraction is done on an offline machine to ensure the security of the LAPD video. Subsequent analysis, including the Nystr\"{o}m extension and the graph MBO scheme, is performed on a 2.3GHz machine with Intel Core i7 and 4 GB of memory. Both experiments on the QUAD dataset and the HUJI EgoSeg dataset can be finished within a minute after extracting features; each batch of the LAPD body-worn video dataset (see \cref{subsubsec:lapd_comp} for details) takes around two minutes.  

\begin{table*}
\centering
\caption{Experimental Setup\label{table:para}}
\begin{tabular}{l|rrrr|r|rlrr|rrr}
\toprule
\multicolumn{1}{c}{}& \multicolumn{4}{c}{Motion feature} &\multicolumn{1}{c}{NMF} &\multicolumn{4}{c}{Spectrum of the Graph Laplacian}&\multicolumn{3}{c}{MBO}\\ 
 & \begin{tabular}{@{}c@{}}$\Delta T$ \\ (sec)\end{tabular}& FPS&\begin{tabular}{@{}c@{}}Number of\\ segments\end{tabular} &\begin{tabular}{@{}c@{}}Window size \\ (segment)\end{tabular} & $\hat{k}$ & $N_{eig}$ & $\tau_{ij}$&  \begin{tabular}{@{}c@{}}$N_{sample}$ \\ \end{tabular}&\begin{tabular}{@{}c@{}}Batch size \\ (segment)\end{tabular}& $\eta$ & $\Delta t$ & $N_{step}$\\
\midrule
QUAD &$1/60$&60&14,399&-&50&500&$\tau = 1$&  1000 & - & 300 & 0.1 & 10\\
LAPD&$1/5$&30&274,443&5  & 50 & 2000 &$K = 100$& 2000 & 30000 & 400 & 0.1 & 10\\
LAPD \cite{egomotionclassification} & $1/5$&30&274,443&-&-&2000&$K = 100$& 2000 &30000 &400 & 0.1 &10\\
HUJI&4 &15&36,421&20 & 50 & 400 &$K = 40$& 400 & - & 300 & 0.1 & 10\\
\bottomrule
\end{tabular}
\end{table*}

\subsection{QUAD dataset}
\label{sec:quad}The authors of \cite{kitani2011fast} choreographed and made public the QUAD dataset, which is about four minutes long and filmed at 60 frames per second. The footage was recorded with a head-mounted Go-Pro Camera while the camera-wearer was undergoing nine ego-activities (reported in TABLE \ref{quad}), such as walking, jumping, and climbing up stairs\footnote{The reported categories of ego-activities are the same ones used in \cite{egomotionclassification} and are different from \cite{kitani2011fast}.}. 
The authors of \cite{kitani2011fast} and \cite{egomotionclassification} tested their ego-activity classification methods on this dataset; we follow the same experimental protocol as \cite{egomotionclassification}. Each video ``segment'' is chosen to be an individual frame and we uniformly sample 10\% segments within each category as fidelity in agreement with the protocol employed  in \cite{egomotionclassification}. Such choice of one frame per segment yields 14,399 segments. 

In TABLE \ref{quad}, we report precision within each category and the mean precision, directly averaged over nine classes; the authors of \cite{kitani2011fast} have also reported the mean precision and the authors of \cite{egomotionclassification} provided detailed precision per class. Both our method and the method in \cite{egomotionclassification} use 10\% of the video, sampled uniformly, as fidelity. The method in \cite{kitani2011fast} is unsupervised and the reported mean precision is calculated after matching the discovered ego-activity categories to the ground-truth categories in a way that the best match gives the highest harmonic mean of the precision and recall (i.e. the best F-measure).    
Our result is overall an improvement upon \cite{egomotionclassification} in terms of precision. 

The QUAD dataset only consists of a short choreographed video, in which activities of interest have a relatively balanced proportion, and the challenges we observe in the field datasets are absent. However, the experiment on the choreographed dataset validates the baseline ability of our method in recognizing ego-activities in body-worn videos.  We further test our method and showcase the applicability of our method to datasets consist of multiple videos of different lengths that are not choreographed and recorded in a variety conditions. 
\begin{table}
\centering
\caption{Class proportion and precision of the QUAD dataset\label{quad}}
\begin{tabular}{lcccc}
\toprule
&& \multicolumn{3}{c}{\textbf{Precision}} \\
\textbf{Class} & 
\textbf{Proportion} & 
\textbf{\cite{kitani2011fast}} &
\textbf{\cite{egomotionclassification}} &
\textbf{Ours} \\
\midrule
Jump  & 14.54\% & - & 92.51\% & 99.07\% \\
Stand & 13.74\% & - &  87.90\% & 87.11\%\\
Walk  & 12.75\% & - & 84.52\% & 98.37\%\\
Step  & 12.65\% & - & 93.98\% & 98.54\%\\
Turn Left & 11.25\%  & - & 89.43\% & 96.96\% \\
Turn Right & 10.16\% & - & 92.80\%  & 96.21\%\\
Run & 9.00\% & - & 92.38\% & 96.17\%\\
Look Up & 8.85\%  & - & 80.36\% & 90.02\% \\
Look Down & 7.06\% & - & 84.59\% & 89.00\%\\
\midrule
\textbf{Mean} & 11.11\% & \textbf{95\%} & 88.74\% & 94.49\%\\
\bottomrule
\end{tabular}
\end{table}
\subsection{LAPD BWV dataset}
\label{subsubsec:lapd_comp}
The LAPD body-worn video dataset consists of 100 videos with a total length of 15.25 hours recorded at 30 frames per second. The video footage is recorded by cameras mounted on police officers' chests when they are performing a variety of law enforcement activities. The dataset consists of videos recorded both inside vehicles and outdoors and under a variety of illumination conditions. We manually annotated each frame of all 100 videos with one of 14 class labels. Although we train on and classify video footage in all 14 categories, we exclude five insignificant classes, such as ``exiting car'' and ``obscured camera'', from performance evaluations of the ego-activity recognition algorithms.  We report activity proportions of the selected classes in TABLE \ref{table: bwv} and, for completeness, all 14 classes in TABLE \ref{table: bwv full} of the Appendix. 

We apply the method in \cite{egomotionclassification} with the provided implementation on the LAPD body-worn video dataset. 
\cite{egomotionclassification} computes a feature vector per frame instead of per short video segment, which consists of 6 frames (0.2 seconds). The average of the frame-wise features over a segment is used as the feature vector of the segment. By doing so, the numbers of video segments to classify in both methods are the same. We apply a moving window average operator with a window size of one second (five segments) to our features. The features of \cite{egomotionclassification} inherently incorporate temporal information, so we use the aggregated segment-wise features as they are without further smoothing. 

We divide the 274,443 segments into 9 disjoint batches, each of which consists of approximately 30,000 segments. As  each segment has a duration of $0.2$ seconds, each batch therefore consists of 100-minutes of footage spanning multiple videos. We perform the classification on each batch independently and concatenate the classification results. We note that both our method and the method proposed in \cite{egomotionclassification} make use of the Nystr\"{o}m extension and the MBO scheme described in \cref{sec:eig} and \cref{sec:class} respectively, so they share the same set of parameters. We choose $N_{sample} = 2000$ and $N_{eig} = 2000$ to be the same for both methods for each batch so that they share the same computation cost and both give good performance relatively to other choices of parameters. We have tuned parameters $\eta$ ranging from $0.01$ to $1000$ and found that $\eta = 400$ and $\tau$ selected automatically according to \cite{zelnik2005self} with $K = 100$ work well for both methods.

With regards to sampling fidelity points, we use the same  protocol as the one used in \cite{egomotionclassification} where we uniformly sample 10\% segments within each class. Consequently, we have many more samples of common activities than rare activities.  

\begin{figure}
\centering
{\includegraphics[width = 0.47\textwidth]{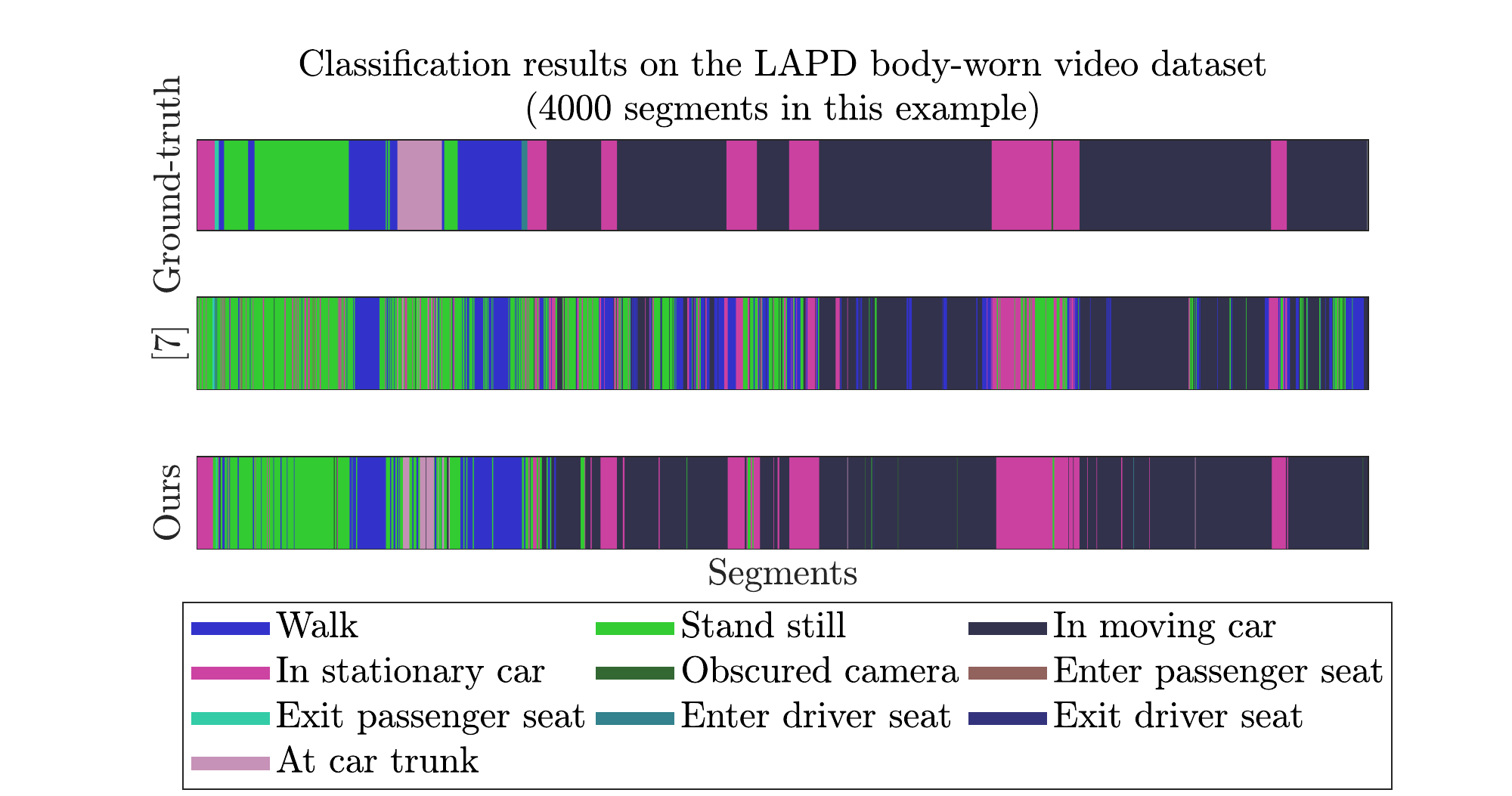}
\caption{Classification results on a contiguous sample of 4000 segments (approximately 13 minutes) from the LAPD body-worn video dataset. The results are obtained by running both methods with the parameters described in \cref{subsubsec:lapd_comp}.
\label{fig:bwv_fig}}
}
\end{figure}

\begin{figure}
\centering{
\subfloat[Method proposed in \cite{egomotionclassification}]{
\includegraphics[width=0.47\textwidth]{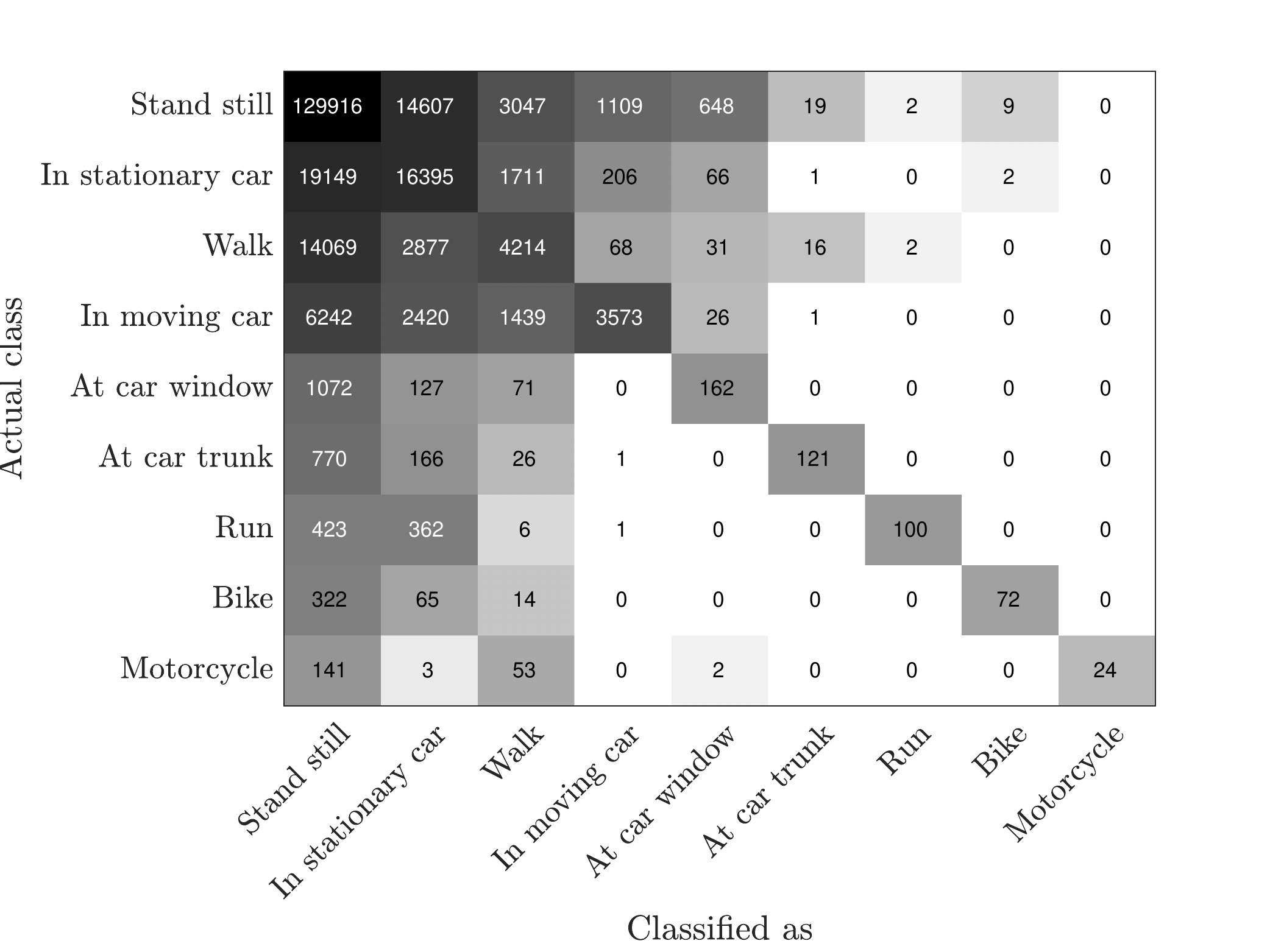}
}\\
\subfloat[Ours]{
\includegraphics[width=0.47\textwidth]{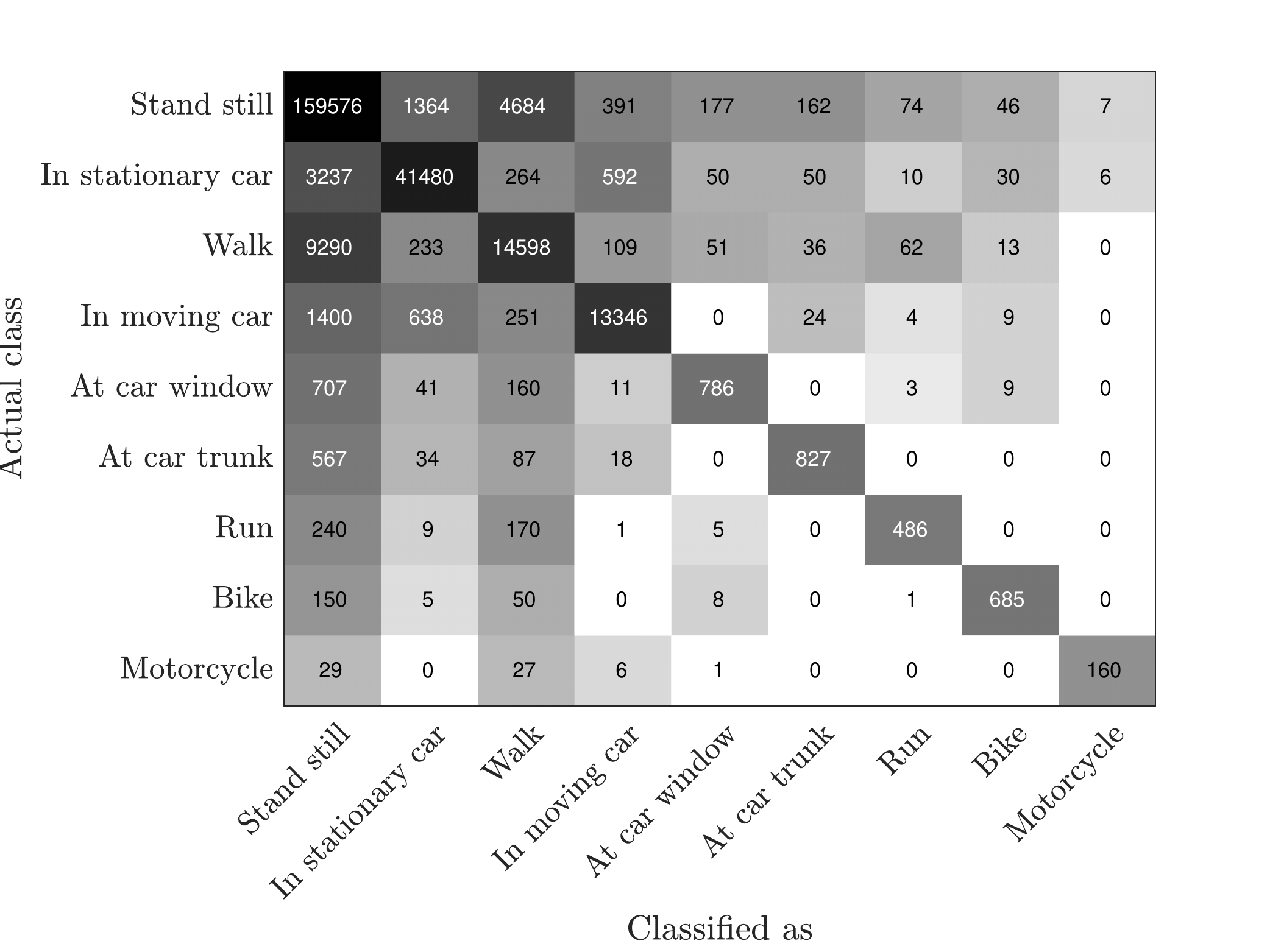}
}
\caption{Confusion matrices for the LAPD Body-worn video dataset. The background intensity in cell $(k,\ell)$ corresponds to the number of data points in class $k$ that are classified as class $\ell$ by the algorithm. \label{fig:cm_bwv}}}
\end{figure}

In TABLE \ref{table: bwv}, we report the precision and recall within each class and their respective means averaged over the selected nine classes. We refer readers to TABLE \ref{table: bwv full} in the Appendix for a full table of all 14 classes as well as the overall accuracy, which is the proportion of video segments that are correctly classified. 
We also present a sample of the color-coded classification results in \cref{fig:bwv_fig} and the confusion matrices in \cref{fig:cm_bwv}. 

Our method outperforms \cite{egomotionclassification} in most of the categories in terms of precision and is a major improvement according to recall. We theorize that the features proposed in \cite{egomotionclassification} are too simple to distinguish among the increased variety of ego-activities in the larger LAPD body-worn video dataset. The features they propose do not make use of the locality of motion within each frame, which we consider crucial in order to differentiate, for instance, driving a car and walking forward. Both activities feature forward motion, but the motion is localized within the windshield region only in the former case. We also note that frequency is a significant component of the features proposed in \cite{egomotionclassification}; however, we do not observe much periodic motion in many ego-activities. 

\begin{table}
\centering
\caption{Class proportion, precision, and recall of the selected nine classes in the LAPD body-worn video dataset \label{table: bwv}}
\begin{tabular}{lccccc}
\toprule
&& \multicolumn{2}{c}{\textbf{Precision}}&\multicolumn{2}{c}{\textbf{Recall}} \\
\textbf{Class} & 
\textbf{Proportion} & 
\textbf{\cite{egomotionclassification}} &
\textbf{Ours} &
\textbf{\cite{egomotionclassification}} &
\textbf{Ours}\\
\midrule
Stand still &  62.57\% & 73.10\% & 89.44\%  &85.42\% &95.24\%\\
In stationary car & 16.84\%  & 41.83\% & 93.69\%  &43.18\%&89.73\%\\
Walk &  9.04\% &38.36\% & 70.53\% & 19.54\% & 59.41\% \\
In moving car& 5.76\%& 70.71\% & 91.03\%& 25.08\%&  84.40\%\\
At car window & 0.64\%& 17.23\%&	71.45\%&10.94\%&45.28\%\\
At car trunk & 0.58\%& 73.78\%&	71.79\%&	11.09\%&	51.78\%\\

Run & 0.33\%&96.15\%&	75.94\%&	11.03\%&	53.35\%\\
Bike & 0.33\% & 85.71\%&	86.49\%&	14.37\%&	75.44\%\\

Motorcycle & 0.08\% &100\%&	92.49\%&	10.76\%&	71.75\%\\
\midrule
\textbf{Mean} & 10.68\% & 66.32\%&  82.54\%& 25.71\%&  69.60\%\\
\bottomrule
\end{tabular}
\end{table}


\subsection{HUJI EgoSeg dataset}
\label{sec:huji}We also evaluate the performance of our method on the HUJI EgoSeg dataset \cite{poleg2014temporal} \cite{poleg2016compact}. This dataset contains 65 hours of egocentric videos including 44 videos shot using a head-mounted GoPro Hero3+, the Disney dataset \cite{fathi2012social} and other YouTube videos\footnote{The HUJI EgoSeg dataset can be downloaded at \protect\url{http://www.vision.huji.ac.il/egoseg/videos/dataset.html}.}. The dataset contains 7 ego-action categories: \textit{Walking, Driving, Riding Bus, Biking, Standing, Sitting, and Static}. We normalize the frame rate of each video to 15 frames per second to match with the normalized frame rate in \cite{poleg2016compact}. We divide each video sequence into segments of 4 seconds ($\Delta  T = 4$ seconds, 60 frames), which also matches the length of each video segment in \cite{poleg2016compact}. The activities present in the HUJI EgoSeg dataset are all relatively long-term activities compared to the LAPD Body-worn video dataset, so using longer video segments reduces the number of data points without the risk of missing short-term activities. With such choice of $\Delta T$, we have 36,421 segments. For the Nystr\"{o}m extension and the MBO scheme, we have found that the combination of $N_{sample} = 400$, $N_{eig} = 400$, $\eta = 300$, and $K = 40$ gives satisfactory results. 

\begin{figure}
\centering
\includegraphics[width=0.47\textwidth]{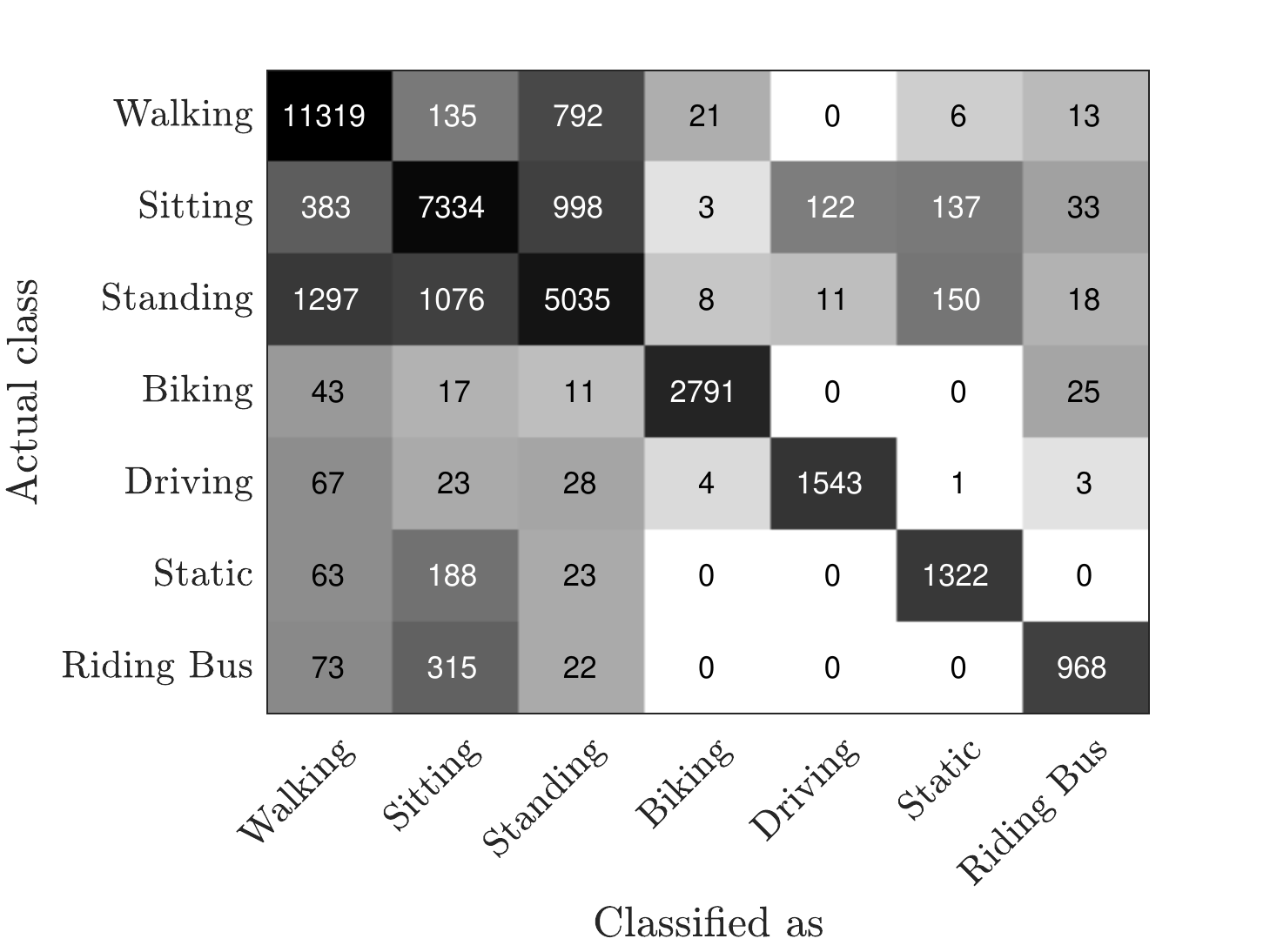}
\caption{Confusion matrix for the HUJI EgoSeg dataset. The background intensity in cell $(k,\ell)$ corresponds to the number of data points in class $k$ that are classified as class $\ell$ by the algorithm. 
\label{cm_huji}}
\end{figure}

\begin{figure}
\centering
{\includegraphics[width = 0.47\textwidth]{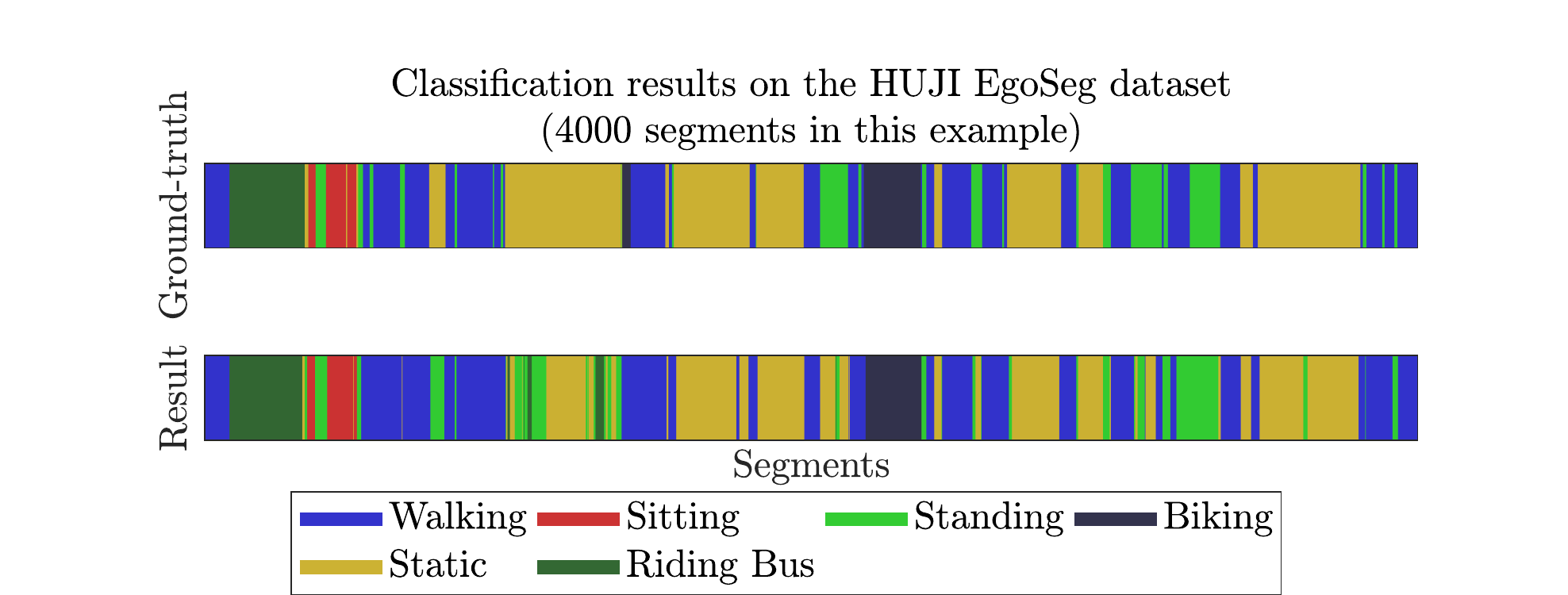}
\caption{Classification results on a contiguous sample of 4000 segments (approximately 4 hours) from the testing set of HUJI EgoSeg dataset. The recall of the same experiment is reported in TABLE \ref{huji}. \label{fig:huji_fig}}
}
\end{figure}

We follow the same experimental protocol of \cite{poleg2014temporal,poleg2016compact} to divide the entire dataset into a training set and a testing set. We randomly pick video sequences until we have 1300 segments (approximately 90 minutes of video) per class as the training set, and we uniformly sample 10\% of the training set as fidelity points, which is about 10\% of the training data used in \cite{poleg2016compact}\footnote{The authors of \cite{poleg2016compact} do not explicitly mention the fidelity percentage; we estimate the percentage according to their released code at \url{http://www.vision.huji.ac.il/egoseg/}.}. In this experiment, we use recall to evaluate the performance since it is the common measure of success in \cite{poleg2014temporal,ryoo2015pooled,tran2015learning,poleg2016compact}. TABLE~\ref{huji} details the classification results on the testing set. The classification performance of methods other than ours are reported in \cite{poleg2016compact}.  
We also report the confusion matrix in \cref{cm_huji} and a color-coded sample of the classification result in \cref{fig:huji_fig}. 

We observe that the recalls of Sitting, Standing, and Riding Bus are typically lower than other activities across all five methods, so we believe that these activities are inherently difficult to recognize with motion-based features. According to TABLE \ref{huji}, our method outperforms --- using recall as a measure of success --- other methods that use handcrafted motion and/or appearance features with or without deep convolution neural networks,  with the exception of \cite{poleg2016compact}. We emphasize that our method uses a fraction of the training data of the supervised methods and still achieves comparable results. When we use the entire training set as fidelity, the mean recall only sees a slight increase. 

\begin{table}
{\centering
\caption{Class proportion and recall of the HUJI EgoSeg dataset \label{huji}}
\begin{tabular}{lcccccc}
\toprule
&& \multicolumn{5}{c}{\textbf{Recall}} \\
\textbf{Class} & 
\textbf{Proportion} & 
\textbf{\cite{poleg2014temporal}} &
\textbf{\cite{ryoo2015pooled}} &
\textbf{\cite{tran2015learning}} &
\textbf{\cite{poleg2016compact}} &
\textbf{Ours} \\
\midrule
Walking & 34\% & 83\% & 91\% & 79\% & 89\% & 91\% \\
Sitting & 25\% & 62\% & 70\% & 62\%  & 84\% & 71\%\\
Standing &21\% & 47\% & 44\% & 62\% & 79\% & 47\%\\
Biking & 8\% & 86\% & 34\% & 36\% & 91\% & 88\% \\
Driving & 5\% & 74\% & 82\% & 92\% & 100\%& 95\%\\
Static & 4\% & 97\% & 61\% & 100\% & 98\% & 96\% \\
Riding Bus & 4\% & 43\% & 37\% & 58\% & 82\% & 84\%\\
\midrule
\textbf{Mean} & 14\% & 70\% & 60\% & 70\% & 89\% & 82\%\\
\textbf{Training}&& $\sim$60\%& $\sim$60\%& $\sim$60\%& $\sim$60\%& 6\%\\
\bottomrule
\end{tabular}
}
\end{table}

\section{Conclusion and Future Work}
\label{sec:conclusion}
In this paper, we study ego-activity recognition in first-person video with an emphasis on the application to real-world police body-worn video.  We propose a system for classifying ego-activities in body-worn video footage using handcrafted features and a graph-based semi-supervised learning method. These features based on motion cues do not identify people or objects in the scene and hence secure any personally identifiable information within the video. Our experiments also illustrate that the features are able to differentiate a variety of ego-activities and yield better classification results than an earlier work \cite{egomotionclassification} with the same graph-based semi-supervised learning method. The semi-supervised classification method addresses the challenge of insufficient training data; it achieves comparable performance to supervised methods on two publicly available benchmark datasets using only a fraction of training data. Despite using a smaller fraction of training data, our classification results are comparable to or better than those in prior works, which include both classical and deep-learning methodologies.  The proposed system also demonstrates promising results on field data from body-worn cameras used by the Los Angeles Police Department.  

We note that the MBO-based classification method can be used with any feature design, not only the global motion descriptor as presented here. The general graphical setting of the classification method even allows features that cannot be represented by a vector in the Euclidean space so long there exists a way to measure similarity between the features of two data points. Also, the Nystr\"{o}m extension is still applicable with new features or similarity measures to efficiently approximate the spectrum of the graph Laplacian. 

Recent developments in unsupervised convolution neural networks \cite{bhatnagarunsupervised,wang2018deep} might be used to improve and extend the current feature selection method, although caution must be taken to prevent the recovery of personal identifiable information from the learned features. Better incorporating temporal information is another way to move forward. We observe that the police body-worn video dataset contains a mix of long-term and short-term activities, making it difficult to select a single time scale to design the features around. In the present experiment, we chose the length of each segment to be $0.2$ seconds in order to capture short-term activities, but this was redundant for recognizing long-term activities. We chose $\Delta T$ to be four seconds for the HUJI EgoSeg dataset, which significantly reduced the computation cost without sacrificing accuracy, but this was only possible because all activities in the HUJI EgoSeg dataset have long durations. Designing features that efficiently handle a mix of long-term and short-term activities is another challenge to be addressed.

Future work will also be directed towards improving the propose classification method. For instance, \cite{jacobs2018auction} recently proposed to incorporate the knowledge of the proportions of classes as an extra input in the semi-supervised classification method described in \cref{sec:class}. Considering the heterogeneity in the class distribution that we observe in the police body-worn video data set (see \cref{table: bwv}), we expect to see an improvement in the classification performance with the class proportion information.

Despite our best effort to develop an accurate system for the classification of police body-worn videos, the variability of the data leads to imperfect classification. Our classification method is naturally paired with uncertainty quantification (UQ)~\cite{bertozzi2018uncertainty}. Besides giving a video segment an ego-activity label, we may use this technology to estimate a measure of uncertainty, which identifies hard-to-classify video segments that require further investigation. Moreover, the measure of uncertainty can suggest footage for police analysts to label to train classification algorithms making an efficient use of human labeling effort. We expect that further development of the feature selection, classification, and uncertainty quantification methodologies will facilitate an implementation of the propose system to be used by law enforcement agencies to summarize a large volume of body-worn video footage.

\section{Acknowledgement}
This work used computational and storage services associated with the Hoffman2 Shared Cluster provided by UCLA Institute for Digital Research and Education’s Research Technology Group. We thank Zhaoyi Meng, Xiyang Luo, and Matt Jacobs for helpful discussion and sharing their code. We also thank the Los Angeles Police Department for providing the body-worn video dataset.

\bibliographystyle{IEEEtran}
\bibliography{bibliography.bib}
\section{Appendix}
\begin{figure}
\centering{
\subfloat[Method proposed in \cite{egomotionclassification}]{
\includegraphics[width=0.47\textwidth]{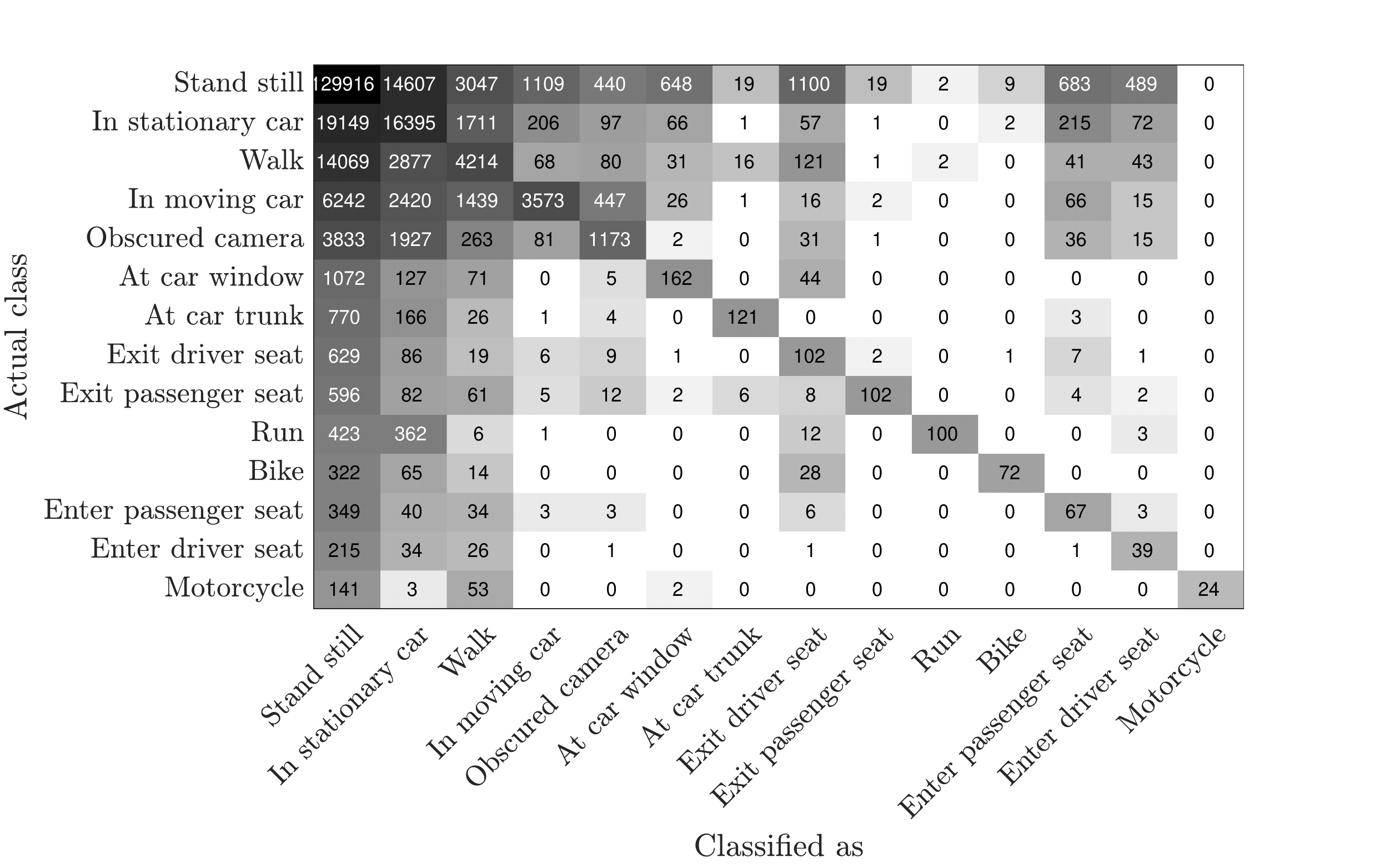}
}\\
\subfloat[Ours]{
\includegraphics[width=0.47\textwidth]{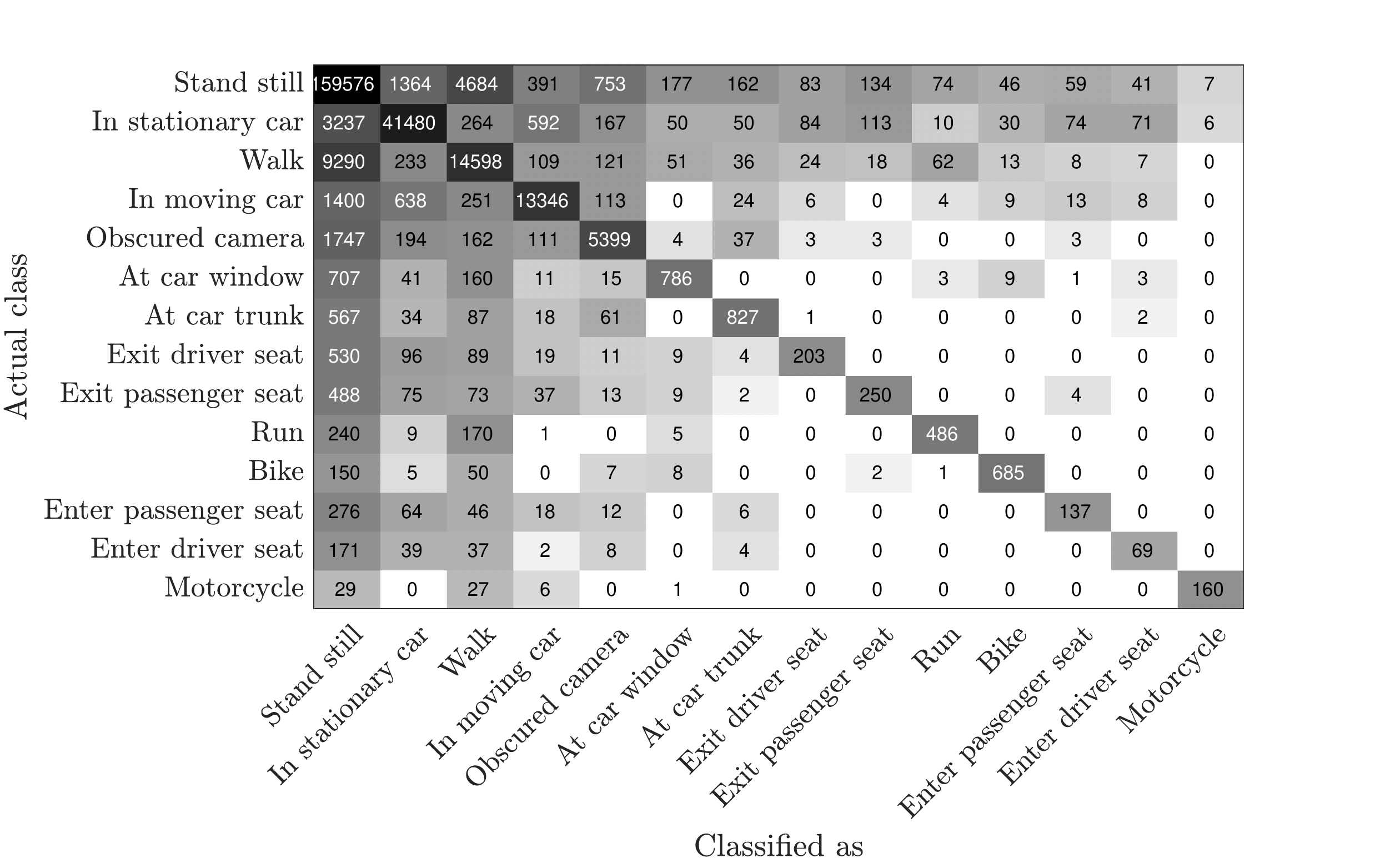}
}
\caption{Confusion matrices for the LAPD police Body-worn video dataset. The background intensity of cell $(k,\ell)$ corresponds to the number of data points in class $k$ that are classified as class $\ell$ by the algorithm. \label{fig:cm_bwv_full}}}
\end{figure}

We report the classification results of the entire 14 ego-activity categories in the LAPD body-worn video dataset in TABLE \ref{table: bwv full} as well as the full confusion matrices in \cref{fig:cm_bwv_full}. 
   
\begin{table}
\centering
\caption{Class proportion, precision , recall, and accuracy on the LAPD body-worn video dataset \label{table: bwv full}}
\begin{tabular}{lccccc}
\toprule
&& \multicolumn{2}{c}{\textbf{Precision}}&\multicolumn{2}{c}{\textbf{Recall}} \\
\textbf{Class} & 
\textbf{Proportion} & 
\textbf{\cite{egomotionclassification}} &
\textbf{Ours} &
\textbf{\cite{egomotionclassification}} &
\textbf{Ours}\\
\midrule
Stand still &  62.57\% & 73.10\% & 89.44\%  &85.42 \% &95.24\%\\
In stationary car & 16.84\%  & 41.83\% & 93.69\%  &43.18\%&89.73\%\\
Walk &  9.04\% &38.36\% & 70.53\% & 19.54\% & 59.41\% \\
In moving car& 5.76\%& 70.71\% & 91.03\%& 25.08\%&  84.40\%\\
Obscured camera & 2.80\%& 51.65\%& 80.82\%&	15.93\%&	70.46\%\\
At car window & 0.64\%& 17.23\%&	71.45\%&10.94\%&45.28\%\\
At car trunk & 0.58\%& 73.78\%&	71.79\%&	11.09\%&	51.78\%\\
Exit driver& 0.35\%&6.68\%&	50.25\%&	11.82\%&	21.12\%\\ 
Exit passenger& 0.34\%& 79.69\%&	48.08\%&	11.59\%&	26.29\%\\
Run & 0.33\%&96.15\%&	75.94\%&	11.03\%&	53.35\%\\
Bike & 0.33\% & 85.71\%&	86.49\%&	14.37\%&	75.44\%\\
Enter passenger& 0.20\%  &5.97\%&	45.82\%&	13.27\%&	24.51\%\\
Enter driver& 0.12\%& 5.72\%&	34.33\%&	12.3\%&	20.91\%\\
Motorcycle & 0.08\% &100\%&	92.49\%&	10.76\%&	71.75\%\\
\midrule
\textbf{Mean} & 7.14\% & 53.33\%&  71.58\%& 21.17\%&  56.41\%\\
\textbf{Accuracy}& &65.03\% & 88.15\% \\
\bottomrule
\end{tabular}
\end{table}
\newpage
\begin{IEEEbiography}[{\includegraphics[width=1in,height=1.25in,clip]{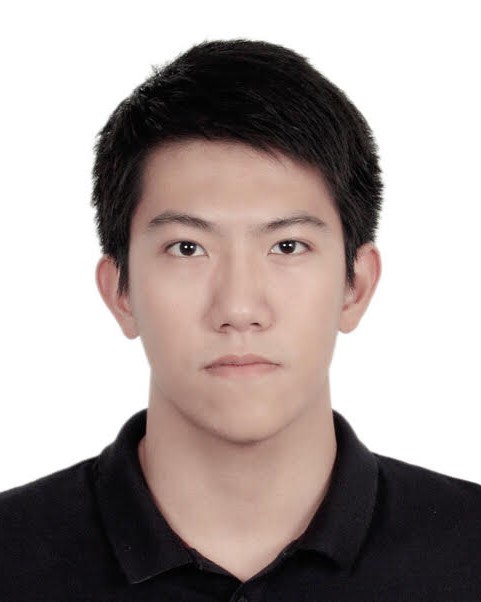}}]{Honglin Chen}
Honglin Chen received the B.S. degree in Mathematics of Computation from the University of California, Los Angeles, CA, USA in 2018.

From 2015 to 2018, he was an undergraduate research assistant with the Computer Vision \& Graphics Laboratory at UCLA and the Center for Brains, Minds and machines (CBMM) at the Massachusetts Institute of Technology under the supervision of Dr. Demetri Terzopoulos and Dr. Tomaso Poggio respectively. His research interest includes human vision, computer vision, computer graphics, machine learning, deep learning, and computational neuroscience.
\end{IEEEbiography}
\vskip 0pt plus -1fil
\begin{IEEEbiography}[{\includegraphics[width=1in,height=1.25in,clip]{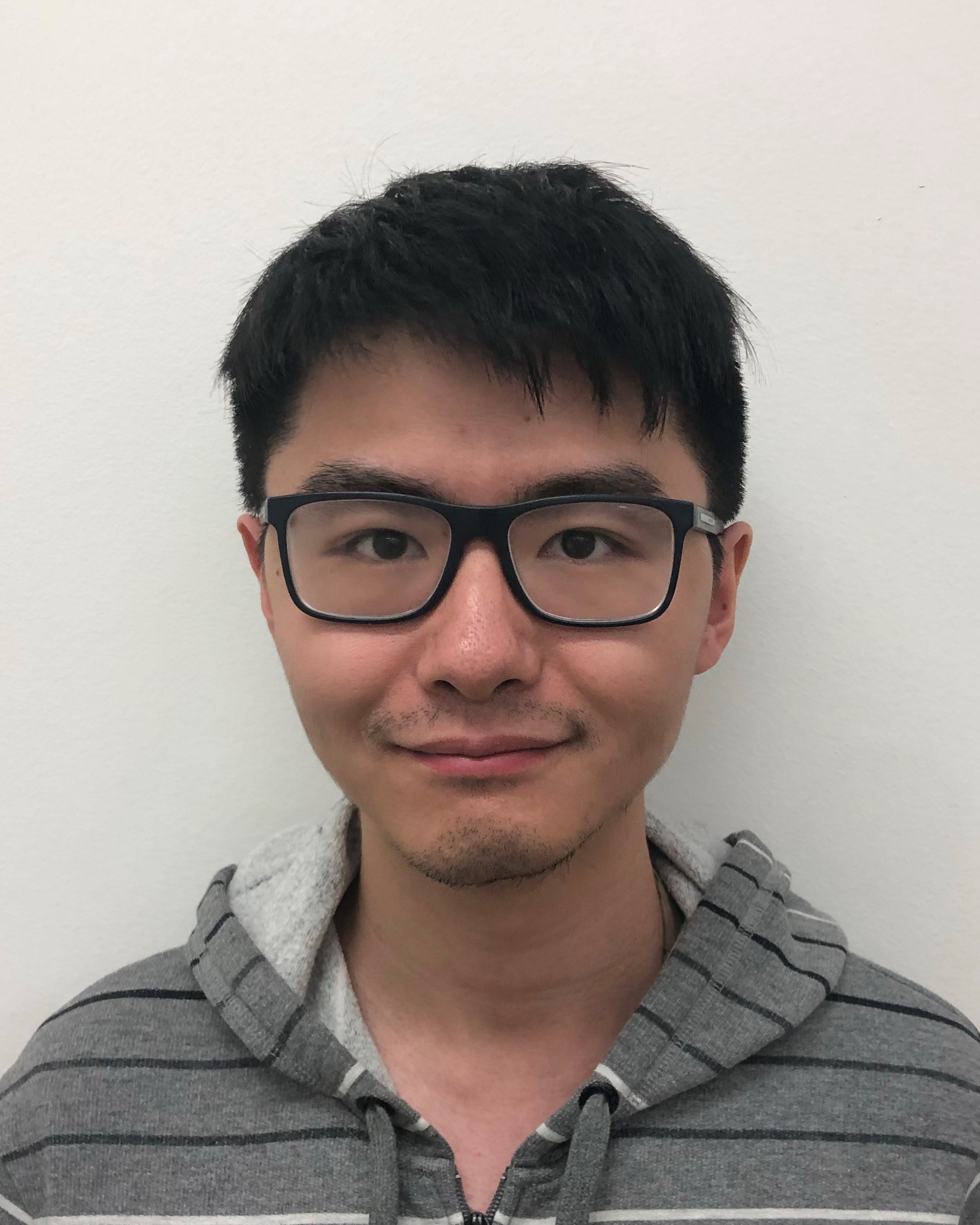}}]{Hao Li}
Hao Li received the B.S. degree in the Mathematics of Computation and the M.A. degree in Applied Mathematics at the University of California, Los Angeles, CA, USA in 2016.
He is currently a third-year graduate student in the Department of Mathematics at UCLA working toward his Ph.D. degree. 

He is a Research Assistant with UCLA since 2016 and graduate student mentor in the UCLA Applied Math REU (Research Experience for
Undergraduates) in the summers from 2016 to 2018. His research interests lie in numerical optimization and Bayesian uncertainty quantification with their applications to data analysis and machine
learning. 
\end{IEEEbiography}
\vskip 0pt plus -1fil
\begin{IEEEbiography}[{\includegraphics[width=1in,height=1.25in,clip]{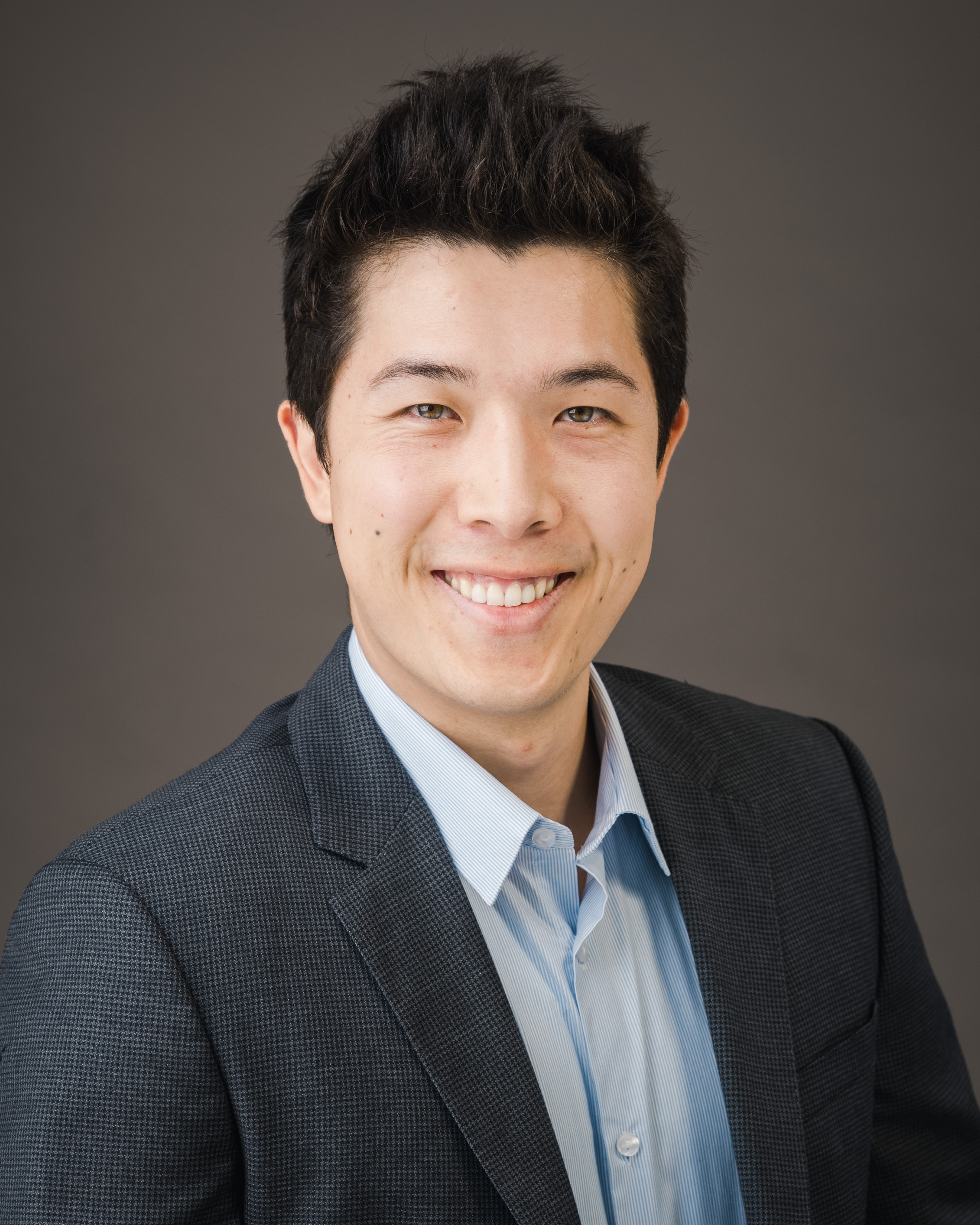}}]{Alexander Song} 
Alexander Song was born in Stockton, CA, USA in 1991. He received a B.A. in philosophy
from the University of California, Berkeley, CA, USA in 2013 and a B.S. in mathematics from
the University of California, Santa Barbara, CA, USA in 2018.

In the summer of 2011, he worked as a Mathematics Tutor and Teaching Assistant for UC
Berkeley’s Summer Bridge Program, a residential academic program for incoming freshmen.
He received a 2014-15 Fulbright grant to teach English and represent the United States as a
“cultural ambassador” in South Korea. He was an Undergraduate Researcher at UC Santa
Barbara’s 2016 Math Summer Research Program, where he investigated structured
linearizations of structured matrix polynomials in collaboration with two other undergraduates
and under the supervision of faculty mentors. The results of this collaboration were published
in the article “Explicit Block-Structures for Block-Symmetric Fiedler-like Pencils” in the
Electronic Journal of Linear Algebra in 2018. In the summer of 2017, he was an
Undergraduate Researcher at the UCLA Applied Math REU (Research Experience for
Undergraduates), conducting research in semi-supervised learning and its application to ego-action classification in body-worn video. He is currently seeking a position in industry as a
data scientist.

He received a Cal Alumni Association Leadership Award, highest honors in general
scholarship upon graduation from UC Berkeley, and membership in Phi Beta Kappa.
\end{IEEEbiography}
\vskip 0pt plus -1fil
\begin{IEEEbiography}[{\includegraphics[width=1in,height=1.25in,clip]{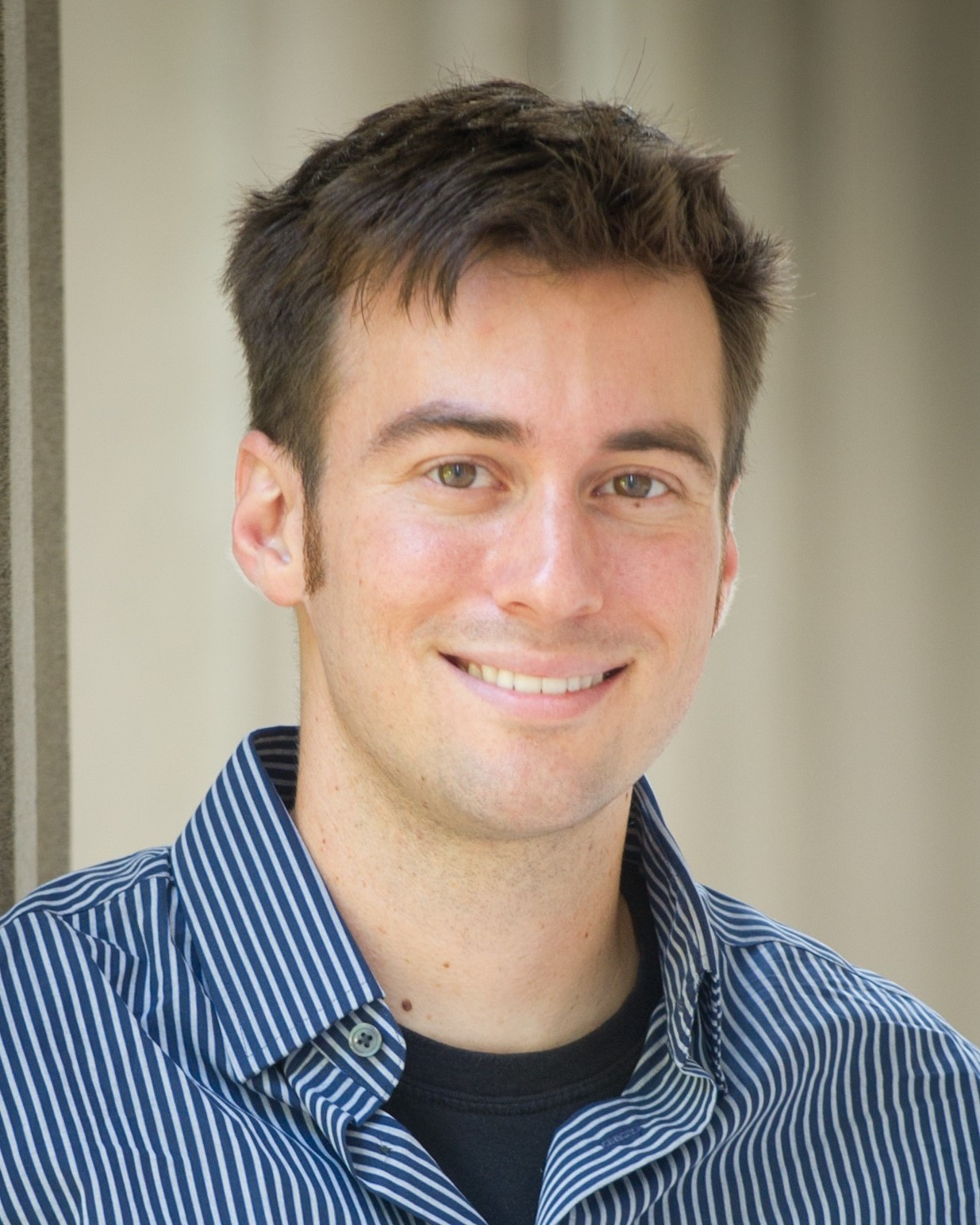}}]{Matt Haberland}
Matt Haberland received the B.S. and M.Eng. degrees in mechanical engineering from Cornell University, Ithaca, NY in 2007 and the Ph.D. degree in mechanical engineering from the Massachusetts Institute of Technology, Cambridge, MA in 2014. 

From 2007 to 2009, he was an engineer at the Jet Propulsion Laboratory in Pasadena, CA, where he created the rock drill contact sensor/stabilizer for the Mars rover Curiosity. From 2014 to 2018, he was an Assistant Adjunct Professor in the Department of Mathematics at the University of California, Los Angeles. Currently, he is an Assistant Professor in the BioResource and Agricultural Engineering Department at the California Polytechnic State University, San Luis Obispo. He has published research in journals including IEEE Robotics and Automation Letters, Bioinspiration and Biomimetics, and Robotica. His research interests include agricultural automation, applications of machine learning, control of legged robots, and linear programming. 
\end{IEEEbiography}
\vskip 0pt plus -1fil
\begin{IEEEbiography}[{\includegraphics[width=1in,height=1.25in,clip]{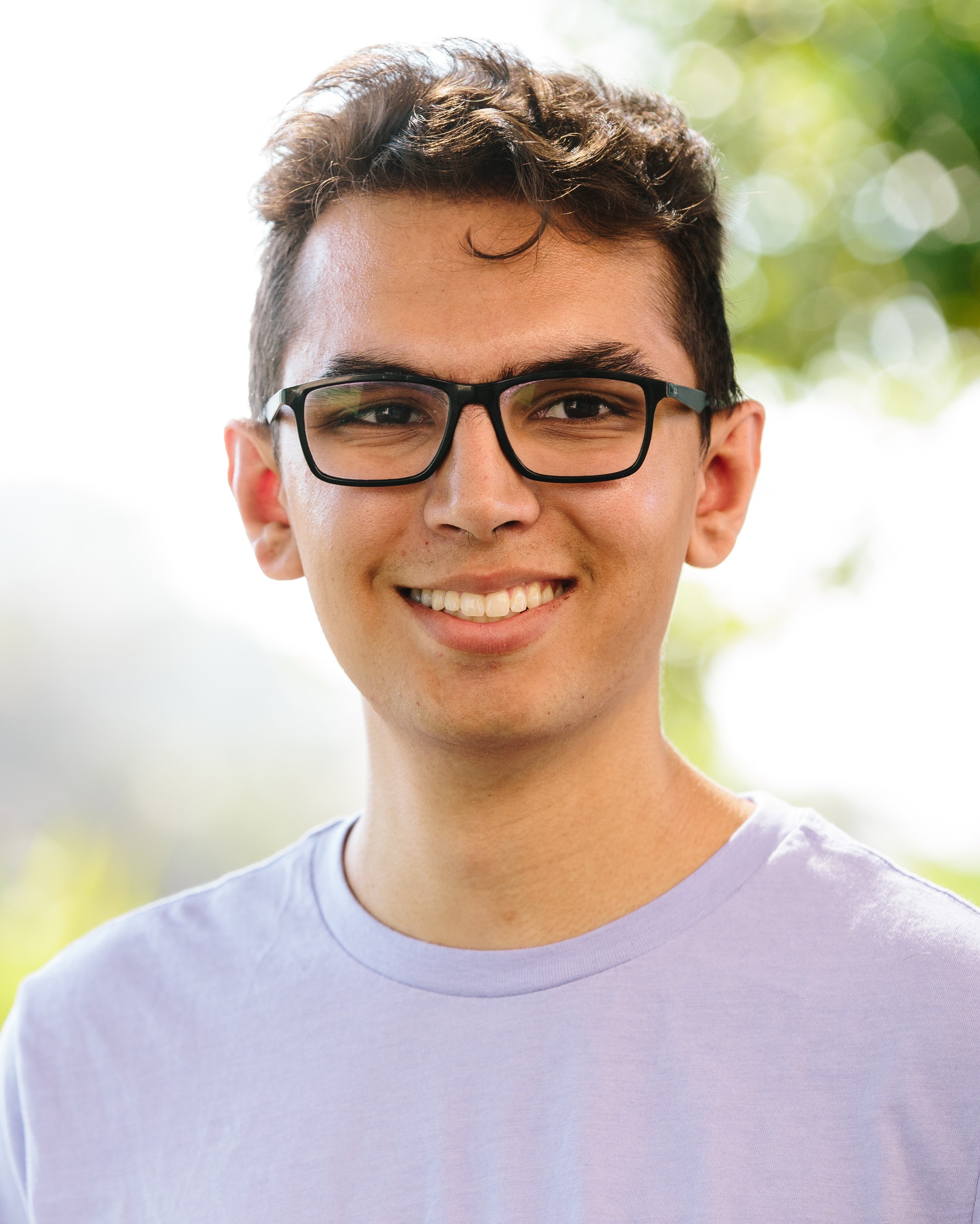}}]{Adam Dhillon} 
Adam Dhillon is a senior undergraduate student at Harvey Mudd College, Claremont, CA, USA. He expects to receive his B.S. in mathematics in May 2019. He will attend a Ph.D. program in mathematics in Fall 2019. 

He is interested in methods of optimizing travel under uncertainty of conditions and approximating optimal trajectories under randomly changing conditions. He has also investigated the effectiveness of statistics on circular data. He recently worked on a notion of complexity based on the number of integrators required to generate a function with an analog computer. 
\end{IEEEbiography}
\vskip 0pt plus -1fil
\begin{IEEEbiography}[{\includegraphics[width=1in,height=1.25in,clip]{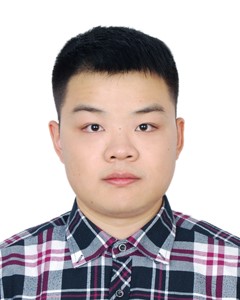}}]{Tiankuang Zhou} 
Tiankuang Zhou received the B.Eng. in Electronic Information Engineering from
University of Science and Technology of China, Hefei, China in 2018. He is currently a Ph.D.
student in Department of Automation in Tsinghua University, Beijing, China. His research
interests include optical computing, computational imaging and machine learning.
\end{IEEEbiography}
\vskip 0pt plus -1fil
\begin{IEEEbiography}[{\includegraphics[width=1in,height=1.25in,clip]{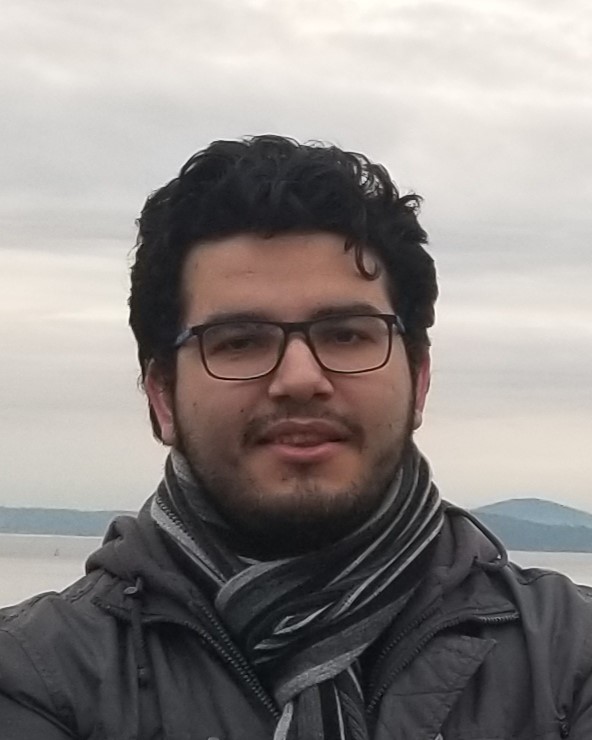}}]{Osman Akar} 
Osman Akar was born in Izmir, Turkey in 1996. He is currently undergraduate student at University of California, Los Angeles. He expects to have his B.S. degree in Mathematics of Computation and M.A. degree in Mathematics by July 2019. He is also prospective Applied Mathematics PhD student of UCLA. This is his second paper, and he expects to study numerical PDEs and image processing in his graduate studies.

He is a recipient of Mathematics Undergraduate Merit Scholarship of UCLA Mathematics department. His awards and honors include IMO Gold Medal and William L. Putnam Competition Honorable Mention.
\end{IEEEbiography}
\vskip 0pt plus -1fil
\begin{IEEEbiography}[{\includegraphics[width=1in,height=1.25in,clip]{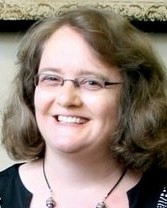}}]{Andrea L. Bertozzi} 
Andrea L. Bertozzi received the B.A., M.A., and Ph.D. degrees in mathematics from Princeton University, Princeton, NJ, USA, in 1987, 1988, and 1991, respectively. 

She was on the faculty of the University of Chicago, Chicago, IL, USA, from 1991 to 1995 and Duke University, Durham, NC, USA, from 1995 to 2004. During 1995–1996, she was the Maria Goeppert-Mayer Distinguished Scholar with Argonne National Laboratory. Since 2003, she has been a Professor of mathematics with the University of California, Los Angeles, where he currently serves as the Director of Applied Mathematics. 

In 2012, she was appointed the Betsy Wood Knapp Chair for Innovation and Creativity. Her research interests include machine learning, image processing, cooperative control of robotic vehicles, and swarming. Dr. Bertozzi is a Fellow of the Society for Industrial and Applied Mathematics, the American Mathematical Society, and the American Physical Society. She is a member of the US National Academy of Sciences and a fellow of the American Academy of Arts and Sciences.
\end{IEEEbiography}
\vskip 0pt plus -1fil
\begin{IEEEbiography}[{\includegraphics[width=1in,height=1.25in,clip]{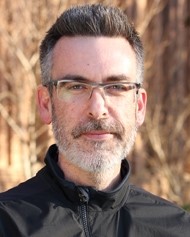}}]{P. Jeffrey Brantingham} 
P. Jeffrey Brantingham received the B.A. in Anthropology from the University of British Columnbia, Vancouver, Canada in 1992, the M.A. and Ph.D. in Anthropology from University of Arizona, Tucson, AZ, USA, in 1996 and 1999 respectively. 

During 2000-2002, he was a postdoctoral scholar at Santa Fe Institute, NM, USA. Since 2002, he has been a Professor of Anthropology with the University of California, Los Angeles, CA, USA.

During 2010-2012, he Served on the LAPD Community-Police Advisory Boards for Counter-Terrorism. He is a co-founder of PredPol, a company that delivers real-time predictive policing to law enforce agencies.
\end{IEEEbiography}
\end{document}